\documentclass[11pt,a4paper]{article}
\pdfoutput=1

\usepackage{jcappub}
\usepackage{amsmath,amssymb,color,bm, graphicx}
\usepackage{enumerate,xstring, hyperref}
\def\be{\begin{equation}}
\def\ee{\end{equation}}
\def\bea{\begin{eqnarray}}
\def\eea{\end{eqnarray}}
\newcommand{\vs}{\nonumber\\}
\def\ba#1\ea{\begin{align}#1\end{align}}
\def\bg#1\eg{\begin{gather}#1\end{gather}}

\newcommand{\fkonenl}{k\frac{P^\prime_m(k)}{P_m(k)}}

\newcommand{\lcdm}{\Lambda{\rm CDM}}

\def\knl{k_\text{NL}}

\newcommand{\refeq}[1]{Eq.~(\ref{eq:#1})}

\newcommand{\reffig}[1]{Fig.~\ref{fig:#1}}

\def\Plin{P_{\rm L}}
\def\Pnl{P_m}

\renewcommand{\v}[1]{\bm{#1}}
\newcommand{\vx}{\v{x}}

\newcommand{\vk}{\v{k}}

\newcommand{\vp}{\v{p}}
\newcommand{\vn}{\v{n}}

\newcommand{\vtheta}{\v{\theta}}
\newcommand{\vell}{\v{\ell}}
\newcommand{\<}{\langle}
\renewcommand{\>}{\rangle}

\renewcommand{\d}{\delta}

\newcommand{\ssca}{\delta\delta}
\newcommand{\sscb}{KK}
\newcommand{\sscc}{\delta K}
\newcommand{\sscd}{K \delta}

\def\be{\begin{equation}}
\def\ee{\end{equation}}
\def\ben{\begin{eqnarray}}
\def\een{\end{eqnarray}}
\def\ba{\begin{array}}
\def\ea{\end{array}}

\def\ba#1\ea{\begin{align}#1\end{align}}

\newcommand{\bq}{\begin{eqnarray}}
\newcommand{\eq}{\end{eqnarray}}
\newcommand{\bes}{\begin{subequations}}
\newcommand{\ees}{\end{subequations}}

\def\R{\mathcal{R}}

\def\P{\mathcal{P}}
\def\W{\mathcal{W}}

\def\T{\mathcal{T}}
\def\fii{F_2}
\def\fiii{F_3}
\def\kunit{\:h\,{\rm Mpc}^{-1}}

\makeatletter
\newlength{\apb@width}
\newcommand{\autoparbox}[2][c]{\settowidth{\apb@width}{#2}\parbox[#1]{\apb@width}{#2}}
\newcommand{\includegraphicsbox}[2][]{\autoparbox{\includegraphics[#1]{#2}}}
\makeatother

\DeclareMathOperator{\cov}{Cov}
\DeclareMathOperator{\bfcov}{\bf Cov}
\DeclareMathOperator{\covkappa}{Cov_{\kappa}}

\newcommand{\comment}[1]{}

\begin{document}

\title{Complete super-sample lensing covariance in the response approach}

\author{Alexandre Barreira,$^1$}
\emailAdd{barreira@MPA-Garching.MPG.DE}
\affiliation{$^1$Max-Planck-Institut f{\"u}r Astrophysik, Karl-Schwarzschild-Str.~1, 85741 Garching, Germany}

\author{Elisabeth Krause,$^{2,3}$ and}
\emailAdd{ekrause@caltech.edu}
\affiliation{$^2$California Institute of Technology, 1200 E California Blvd, Pasadena, CA 91125, U.S.A.\newline
$^3$Jet Propulsion Laboratory, 4800 Oak Grove Dr, Pasadena, CA 91109, U.S.A.}

\author{Fabian Schmidt$^1$}
\emailAdd{fabians@MPA-Garching.MPG.DE}

\abstract{We derive the complete super-sample covariance (SSC) of the matter and weak lensing convergence power spectra using the power spectrum response formalism to accurately describe the coupling of super- to sub-survey modes. The SSC term is completely characterized by the survey window function, the nonlinear matter power spectrum and the full first-order nonlinear power spectrum response function, which describes the response to super-survey density and tidal field perturbations. Generalized separate universe simulations can efficiently measure these responses in the nonlinear regime of structure formation, which is necessary for lensing applications. We derive the lensing SSC formulae for two cases: one under the Limber and flat-sky approximations, and a more general one that goes beyond the Limber approximation in the super-survey mode and is valid for curved sky applications. Quantitatively, we find that for sky fractions $f_{\rm sky} \approx 0.3$ and a single source redshift at $z_S=1$, the use of the flat-sky and Limber approximation underestimates the total SSC contribution by $\approx 10\%$. The contribution from super-survey tidal fields to the lensing SSC, which has not been included in cosmological analyses so far, is shown to represent about $5\%$ of the total lensing covariance on multipoles $\ell_1,\ell_2 \gtrsim 300$. The SSC is the dominant off-diagonal contribution to the total lensing covariance, making it appropriate to include these tidal terms and beyond flat-sky/Limber corrections in cosmic shear analyses.}

%\keywords{}

\date{\today}

\maketitle
\flushbottom

%%%%%%%%%%%%%%%%%%%%%%%%%%%%%%%%%%%%%%%%%%%%%%%%%%%%%%%%%%%%%%%%%%%%%%%%%%%
%%%%%%%%%%%%%%%%%%%%%%%%%%%%%%%%%%%%%%%%%%%%%%%%%%%%%%%%%%%%%%%%%%%%%%%%%%%

\section{Introduction}\label{sec:intro}

One of the main goals of current and future large-scale structure surveys is to constrain the parameters of cosmological models and infer their overall goodness of fit to the data. Central to this exercise is the likelihood function $\mathcal{L}({\bf D}|{\bf M}(\Theta))$, which measures the likelihood that the data vector ${\bf D}$ is a realization of a model characterized by a set of parameters $\Theta$, which makes the prediction ${\bf M}$ for the data. The inference of the model parameters is then given by the posterior distribution, which according to Bayes' theorem is given by $\P(\Theta|{\bf D})  = \mathcal{L}({\bf D}|{\bf M}(\Theta))\P(\Theta)/\P({\bf D})$, where $\P(\Theta)$ and $\P({\bf D})$ are called the prior and evidence, respectively. In most real data analyses, the observed data vector ${\bf D}$ is assumed to be Gaussian distributed, in which case the likelihood is given by a multivariate Gaussian distribution (see e.g.~Refs.~\cite{2016MNRAS.456L.132S, 2017arXiv170704488S} for the limitations of this assumption)
\bq\label{eq:L}
\mathcal{L}({\bf D}|{\bf M}(\Theta)) = \frac{1}{\sqrt{(2\pi)^d{\rm det}(\bfcov)}}{\rm exp}\left[-\frac{1}{2}\left({\bf M}(\Theta) - {\bf D}\right)^{t}\bfcov^{-1}\left({\bf M}(\Theta) - {\bf D}\right)\right]. \nonumber \\
\eq
There are three ingredients in this likelihood function. The first is the data vector ${\bf D}$, e.g., a set of estimated galaxy or lensing power spectra in $d$ wavenumber bins. Second, we have ${\bf M}(\Theta)$, which corresponds to the theoretical model prediction for the ensemble average of the data vector. The final and third piece $\bfcov$ is the covariance matrix, which describes the statistical and systematic errors of the analysis. In this paper, we focus on the calculation of the covariance matrix, which is far less well understood than the model prediction ${\bf M}(\Theta)$, despite its crucial importance for cosmological parameter inference.

The covariance of the three-dimensional matter power spectrum ($\cov(\vk_1, \vk_2)$, or matter covariance for short) is the starting point to the evaluation of the covariance of galaxy or lensing two-point statistics, which are what is actually used in real data analyses (e.g.~Refs.~\cite{2017MNRAS.470.2617A, 2017MNRAS.467.2085G, 2017MNRAS.464.1640S, 2017MNRAS.465.1454H, 2017arXiv170706627J, 2017arXiv170605004V, 2017arXiv170609359K, 2017arXiv170801530D}). Given an estimator $\hat{P}_m(\vk)$ of the matter power spectrum in a wavemode bin centered at $\vk$, the covariance is defined as
\bq
\cov(\vk_1, \vk_2)&=& \langle\hat{P}(\vk_1)\hat{P}(\vk_2)\rangle - \langle\hat{P}(\vk_1)\rangle \langle\hat{P}(\vk_2)\rangle \vs
&=& \cov^\text{G}(\vk_1, \vk_2) + \cov^\text{cNG}(\vk_1, \vk_2) + \cov^\text{SSC}(\vk_1, \vk_2)\,,
\label{eq:covdefintro}
\eq
where the angle brackets denote ensemble averages. The calculation of the matter covariance can be broken down into the evaluation of three terms (more details will be provided in the sections below). These are (i) the Gaussian (G) covariance, (ii) the connected non-Gaussian (cNG) covariance, and (iii) the super-sample covariance (SSC). During the linear regime of structure formation, and for Gaussian-distributed initial conditions, the diagonal Gaussian covariance is the only contribution. It is straightforwardly given in terms of the power spectrum and it is therefore well understood. The cNG term is given by a specific configuration of the matter trispectrum (the Fourier transform of the matter four-point correlation function; cf.~Eq.~(\ref{eq:covNG})) that describes the correlations between observed modes that arise as nonlinear structure formation develops at late times \cite{1999ApJ...527....1S, 2001ApJ...554...56C, 2009MNRAS.395.2065T, 2016JCAP...06..052B, 2016PhRvD..93l3505B, responses1, responses2}. Finally, the SSC term accounts for the correlation between observed modes and modes whose wavelength is larger than the survey size \cite{takada/hu:2013, li/hu/takada, 2014PhRvD..90j3530L, 2014PhRvD..90b3003M, 2014MNRAS.441.2456T, 2014MNRAS.444.3473T}. For instance, if the observed region is embedded in a large-scale super-survey overdensity, then structures in the survey have evolved faster compared to the case where the survey is embedded in a region at cosmic mean density; the SSC term describes this uncertainty (the amplitude of the super-survey overdensity could alternatively be regarded as a signal to be fitted for \cite{2014PhRvD..90j3530L}). We note, for completeness, that in addition to these three sample covariance terms, real analyses must also take into account noise in the data as well as systematic errors of the measurement process.

In Ref.~\cite{responses2}, the authors have used the power spectrum response formalism \cite{responses1} to calculate the cNG term. The power spectrum response functions \cite{wagner/etal:2014, li/hu/takada:2016, 2016JCAP...09..007B, response, 2017arXiv170103375C} describe how the local power spectrum changes as a function of the amplitude of large-scale density and tidal field perturbations. They readily give the squeezed-limit of $N$-point matter correlation functions, and as a result, can be used to describe the coupling between soft (or long-wavelength) and hard (or short-wavelength) modes in perturbation theory \cite{Bernardeau/etal:2002}. The shape of these response functions can be evaluated efficiently in the nonlinear regime of structure formation using separate universe simulations \cite{li/hu/takada, 2014PhRvD..90j3530L, wagner/etal:2014, CFCpaper2, li/hu/takada:2016, lazeyras/etal, response, andreas}, thus permitting to describe these soft-to-hard mode couplings in the nonlinear regime of the hard modes (see Ref.~\cite{responses1} for a more formal introduction to the response formalism in large-scale structure). By working up to 1-loop level in perturbation theory, the authors of Ref.~\cite{responses2} demonstrated that the responses capture the total contribution to the cNG covariance when $k_2 \ll k_1$, with $k_2 < \knl$ and any nonlinear value of $k_1$ (where $k_i$ denotes the amplitude of the vector mode $\vk_i$), as argued before in Ref.~\cite{bertolini1}. For general (non-squeezed) configurations of the modes $k_1, k_2$, the response-approach calculation is able to reproduce the cNG covariance estimated from large ensembles of simulations (e.g.~Ref.~\cite{blot2015}) to within $30\%-40\%$. The usefulness of the response approach in calculations of the matter covariance can ultimately be traced back to the fact that the covariance is dominated by soft-to-hard mode-coupling terms, which are precisely what can be described fully nonlinearly with responses.

In this paper, we turn our attention to the calculation of the SSC term in the response approach. Since it corresponds to the coupling between small-scale modes inside the survey and long wavelength modes outside, it is amenable to be captured completely by response functions. Indeed, in Ref.~\cite{takada/hu:2013}, the authors invoked the so-called {\it trispectrum consistency relation} to derive the SSC as the response of the matter power spectrum to the presence of an infinitely long-wavelength isotropic density perturbation. In previous literature, essentially the same contribution was referred to as {\it beat coupling} \cite{2006MNRAS.371.1188H, 2006PhRvD..74b3522S, 2009MNRAS.395.2065T, 2012JCAP...04..019D} and {\it halo sample variance} \cite{2007NJPh....9..446T, 2013MNRAS.429..344K, 2009ApJ...701..945S, 2003ApJ...584..702H}. The SSC term derived in Ref.~\cite{takada/hu:2013} (and subsequently studied in Refs.~\cite{li/hu/takada, 2014PhRvD..90j3530L} using separate universe simulations) has been incorporated in the analysis of recent galaxy and lensing surveys (see e.g.~Ref.~\cite{2017arXiv170609359K} for a description of the statistical analysis of the DES year 1 results \cite{2017arXiv170801530D}).

Here, we will use the response formalism to rigorously derive and understand the origin of the SSC contributions to the covariance, including tidal contributions not considered in Ref.~\cite{takada/hu:2013}. Recently, Refs.~\cite{akitsu/takada/li, 2017arXiv171100012A, 2017arXiv171100018L} have shown that super-survey tidal fields induce anisotropies (in addition to those induced by redshift-space distortions and the Alcock-Paczy\'nski effect) in the clustering pattern of galaxies. Their derivation made use of a perturbative approach, using the leading-order expressions for the response functions, which is valid only on quasi-linear scales. Here, we focus instead on the impact of the large-scale tidal fields on lensing observables, in particular, on their contribution to the SSC of the two-dimensional monopole lensing convergence power spectrum. We present two derivations of the lensing SSC contribution: one that assumes flat-sky and the Limber approximation for all modes involved, and another that goes beyond the Limber approximation in the super-survey density and tidal perturbations and is valid also for any shape and size of the lensing survey footprint on the sky. By using the generalized separate universe simulation measurements of the density \cite{li/hu/takada, response} and tidal \cite{andreas} matter power spectrum responses, our treatment is valid in the fully nonlinear regime of structure formation. We shall see that the new anisotropic (or tidal) SSC terms can amount to $5\%$ of the total lensing covariance on nonlinear scales. We also demonstrate that, for the survey sky fractions $f_{\rm sky} \approx 0.3-0.4$ that will be attained by future lensing surveys, the flat-sky/Limber assumptions underpredict the SSC contribution (which is the dominant off-diagonal contribution to the total covariance) by about $10\%$. Both tidal and beyond flat-sky/Limber contributions to SSC are not only relevant but also easily evaluated, and should thus be included in cosmological parameter inference analyses using weak lensing data.

The outline of this paper is as follows. In Sec.~\ref{sec:responses}, we begin by reviewing the basics of the response approach, a specific extension of perturbation theory. In Sec.~\ref{sec:covdec}, we define the three-dimensional matter covariance in the presence of a finite survey window function and, in Sec.~\ref{sec:ssc}, we then explicitly use the response approach to derive the full SSC term, which is one of the main results of this paper. In Sec.~\ref{sec:lensing}, we present our two derivations of the SSC covariance of the lensing convergence power spectrum: in Sec.~\ref{sec:lenscovderiv_1}, the derivation based on the flat-sky and Limber approximations for all modes, and in Sec.~\ref{sec:lenscovderiv_2}, a derivation that is valid on the curved sky and goes beyond the Limber approximation in the super-survey modes. We present numerical results for idealized lensing surveys in Sec.~\ref{sec:results}. We summarize and conclude in Sec.~\ref{sec:conc}. In Appendix \ref{app:covderiv}, we present details about the derivation of the matter covariance for finite surveys; Appendix \ref{app:ssctree} provides more intuition about the derivation of the SSC by working at tree-level in perturbation theory; in Appendix \ref{app:limber}, we derive the relation between the spectra and trispectra of the two- and three-dimensional density field under the Limber and flat-sky approximations; and in Appendix \ref{app:general}, we comment on the challenges of going beyond the flat-sky limit and Limber's approximation for all modes in the connected non-Gaussian term.

\section{Power spectrum responses}\label{sec:responses}

In this section, we briefly outline the definition of power spectrum responses. We will limit ourselves to laying down the relevant equations and definitions that will be used throughout, and refer the interested reader to Ref.~\cite{responses1} for a more formal derivation. We adopt the diagram rules of cosmological perturbation theory with the conventions listed in Appendix A of Ref.~\cite{responses1}. Further, in our notation $\vk_{12\cdots n} = \vk_1 + \vk_2 + \cdots + \vk_n$ and $k = |\vk|$ denotes the amplitude of vector $\vk$.

The $n$-th order matter power spectrum response $\R_n$ can be defined with the following interaction vertex
\ba
&\lim_{\{p_a\} \to 0} \left(
\raisebox{-0.0cm}{\includegraphicsbox[scale=0.8]{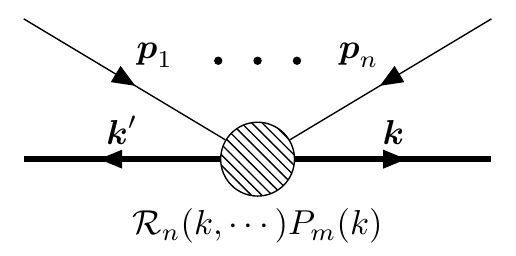}}
\right)   \nonumber \\
\nonumber \\
& = \frac12 \R_n(k;\, \{\mu_{\vk,\vp_a}\},\, \{\mu_{\vp_a,\vp_b}\},\, \{p_a/p_b\}) \Pnl(k) (2\pi)^3 \d_D(\vk+\vk'- \vp_{1\cdots n})\,.
\label{eq:Rndef}
\ea
Physically, it is interpreted as the {\it response} of the nonlinear power spectrum $P_m(k)$ of the small-scale (hard) mode $\vk$ to the presence of $n$ long-wavelength (soft) modes $\vp_1, ..., \vp_n$. The dashed blob thus describes the fully evolved nonlinear matter power spectrum $\Pnl(k)$, as well as all its possible interactions (including loops) with the $n$ long wavelength perturbations. In our notation, $\lim_{\{p_a\} \to 0}$ means that we keep the leading contribution in the limit in which all soft momenta approach zero. The response $\R_n$ depends on the scale $k$, as well as on the cosine of the angles between the soft modes involved and their angles with the hard mode $\vk$; the response does not depend on the absolute value of the soft momenta, but depends on their ratios in general.

With the aid of the diagrammatic representation of the responses, we can establish a link between $\R_n$ and the squeezed limit of the $(n+2)$-point matter correlation function. Explicitly, if we {\it attach} propagators (i.e., power spectra) to the soft momentum lines in Eq.~(\ref{eq:Rndef}), then we can write
\ba
\lim_{\{p_a\}\to 0} \left(
\raisebox{-0.0cm}{\includegraphicsbox[scale=0.8]{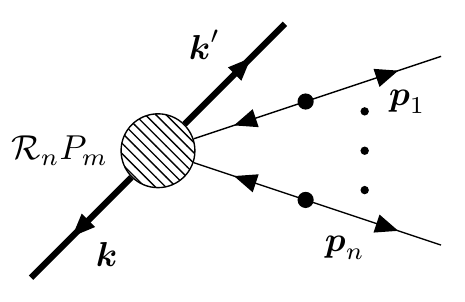}}
+ (\text{perm.}) \right)& = \<\d(\vk)\d(\vk')\d(\vp_1)\cdots \d(\vp_n)\>_{c, \R_n}
\vs
= n!\, \R_n(k;\, \{\mu_{\vk,\vp_a}\},\, \{\mu_{\vp_a,\vp_b}\},\, \{p_a/p_b\})
\Pnl(k) &\left[\prod_{a=1}^n\Plin(p_a)\right] \: (2\pi)^3 \d_D(\vk+\vk'+\vp_{1\cdots n})\,,
\label{eq:sqnpt}
\ea
with the subscript ${}_c$ denoting connected correlators and the $n!$ factor accounting for the permutations of the $\vp_a$. The subscript ${}_{\R_n}$ in the $(n+2)$-connected correlator serves to indicate that only certain contributions to the correlation function are actually captured by $\R_n$. The remaining contributions to $\<\d(\vk)\d(\vk')\d(\vp_1)\cdots \d(\vp_n)\>_c$ are either small in the squeezed limit, or are response-type terms as well, but described by lower order responses $\R_m$, $1\leq m < n$, in conjunction with perturbation theory kernels involving only the soft modes $\vp_a$.

By interpreting the local nonlinear matter power spectrum as a biased tracer of large-scale structure, the  $\R_n$ can be expanded in terms of all local gravitational observables (or operators $O$) associated with the $n$ long-wavelength modes. These operators form a basis $\mathcal{K}_O$ that does not depend on the mode $k$ and that unequivocally specifies the angular structure of $\R_n$:
\be
\R_n(k;\, \{\mu_{\vk,\vp_a}\},\, \{\mu_{\vp_a,\vp_b}\},\, \{p_a/p_b\})
= \sum_O R_O(k) \mathcal{K}_O^{(n)}(\{\mu_{\vk,\vp_a}\},\, \{\mu_{\vp_a,\vp_b}\},\, \{p_a/p_b\})\,.
\label{eq:Rndecomp}
\ee
The functions $R_O(k)$ are called \emph{response coefficients} and their physical interpretation is that they describe the response of the power spectrum to the specific configuration of the large-scale perturbations that corresponds to the operator $O$. At tree level in perturbation theory, the $k$-dependence of the coefficients can be derived analytically by plugging Eq.~(\ref{eq:Rndecomp}) into Eq.~(\ref{eq:sqnpt}). In the nonlinear regime of structure formation, these response functions can be evaluated using separate universe simulations \cite{li/hu/takada, 2014PhRvD..90j3530L, wagner/etal:2014, CFCpaper2, li/hu/takada:2016, lazeyras/etal, response}, recently generalized to tidal fields \cite{andreas}, that simulate infinitely long-wavelength perturbations.

\begin{figure}[t!]
        \centering
        \includegraphics[width=\textwidth]{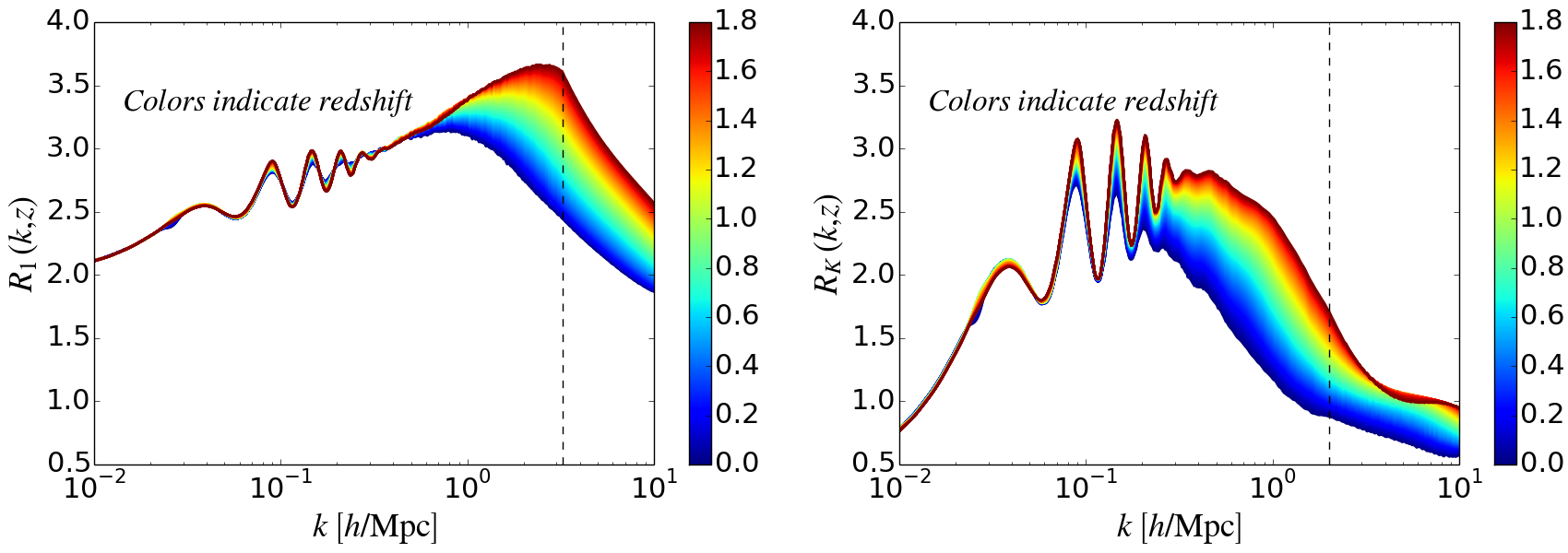}
        \caption{Redshift- and scale-dependence of the $R_1$ (left) and $R_K$ (right) response coefficients (cf.~Eqs.~(\ref{eq:R1dec}), (\ref{eq:R1}) and (\ref{eq:RK})). The curves are color coded by redshift. The result shown is obtained via interpolation from the existing simulation measurements at $z=0,0.5,1,2,3$ for $R_1$ (left; Ref.~\cite{response}) and $z = 0, 0.46, 0.98, 1.89, 2.63$ for $R_K$ (right; Ref.~\cite{andreas}). The vertical dashed lines indicate the maximum value $k_{\rm max}$ up to which $R_1$ and $R_K$ have been measured in Refs.~\cite{response} and \cite{andreas}, respectively; the result for $k > k_{\rm max}$ is obtained with the extrapolation of Eq.~(\ref{eq:G1extra}).}
\label{fig:resp_zevo}
\end{figure}

In this paper, we wish to use responses to evaluate the SSC of the matter power spectrum, for which we shall need only the first order case $n=1$, $\R_1$. Up to first order, there are only two operators with associated kernels $\mathcal{K}_O$, which correspond to a density and a tidal perturbation. Following the notation of Ref.~\cite{responses1} we write
\bq
\label{eq:R1dec} \R_1(k, \mu) &=& R_1(k) + R_K(k)\Big(\mu^2 -\frac13\Big), \\
\label{eq:R1} R_1(k) &=& 1 + G_1(k) - \frac13\fkonenl, \\
\label{eq:RK} R_K(k) &=& G_K(k) - \fkonenl,
\eq
where $\mu \equiv \mu_{\vk,\vp} = \vk\vp/p/k$, $G_1$ and $G_K$ are the so-called density and tidal growth-only response functions, respectively, and a prime denotes a derivative w.r.t.~$k$. The first-order isotropic response coefficient $R_1$ has been measured in the nonlinear regime with separate universe simulations in Ref.~\cite{li/hu/takada}; in Ref.~\cite{response}, the higher-order responses $R_2$ and $R_3$ were measured as well. Simulation measurements of the tidal field first order response $R_K$ have also been recently performed in Ref.~\cite{andreas}. The scale and redshift dependence of these two response coefficients is shown in Fig.~\ref{fig:resp_zevo}. The simulation box setups used in Refs.~\cite{response} and \cite{andreas} ($L_{\rm box} = 500\ {\rm Mpc}/h$ with $N_p = 512^3$ particles) allows for measurements of the responses in the range $k \in \left[k_{\rm fund}, k_{\rm max} \right]$, where $k_{\rm fund} = 2\pi/L_{\rm box} \approx 0.012\ \kunit$ is the fundamental mode and $k_{\rm max}$ is the maximum wavenumber up to which the power spectrum is reliably measured. For the $R_1$ measurements from Ref.~\cite{response} we take $k_{\rm max} = k_{\rm N} = N_p^{1/3} \pi/L_{\rm box} \approx 3.2 \kunit$, where $k_{\rm N}$ is the Nyquist frequency; for the $R_K$ measurements from Ref.~\cite{andreas}, which are based on a fixed-grid particle-mesh simulation, we use $k_{\rm max} = 2 \kunit$. In the lensing calculations we perform below, we will need to evaluate the responses outside this range. For $k < k_{\rm fund}$, we take the linear theory result: $G_1 = 26/21$ and $G_K = 8/7$. For $k > k_{\rm max}$, we extrapolate the growth-only responses as
\bq\label{eq:G1extra}
G_{a}(k > k_{\rm max}) \longrightarrow B_a + \left[G_{a}(k=k_{\rm max}) - B_a\right] \left(\frac{k}{k_{\rm max}}\right)^{-1/2} ,
\eq
where $a \in \{1, K\}$, and $B_a$ are constants to which the $G_{a}$ asymptote for $k \to\infty$. We choose $B_1 \approx 0.75$ and $B_K \approx 2.2$, which, using the slope of the nonlinear power spectrum on very small scales, implies $R_1(k \to\infty) = 1$ and $R_K(k\to\infty) = 0$, which is in accordance with a picture in which, on small scales (where virialization processes inside halos can act to ``erase memory'' from large-scale perturbations) an isotropic density fluctuation contributes only with the so-called reference density effect \cite{response} and a tidal field does not sizeably affect halo density profiles \cite{andreas} (and thus, small scale clustering). The choice of the exponent $-1/2$ was chosen to ensure a smooth transition at $k = k_{\rm max}$. Our conclusions do not depend on the precise form of this high-$k$ extrapolation in any significant way.

\section{Three-dimensional matter power spectrum covariance}\label{sec:cov}

In this section, we derive the expressions of the covariance of the three-dimensional matter power spectrum in the presence of a finite survey window function, which includes the full SSC term. We will follow closely the notation in Sec.~II of Ref.~\cite{takada/hu:2013} to facilitate comparisons.

\subsection{Covariance decomposition}\label{sec:covdec}

Let us define the matter density contrast field measured inside a survey with window function $W(\vx)$ as
\bq\label{eq:dreal}
\delta_W(\vx) = W(\vx) \delta(\vx),
\eq
where $\delta(\vx)$ is the three-dimensional matter density contrast and $W(\vx)$ is unity if $\vx$ is inside the observed region and zero otherwise. The Fourier transform of $\delta_W(\vx)$ is thus given by the convolution of the Fourier transform of the density contrast with the Fourier transform of the window function (tildes indicate Fourier-space quantities)
\bq\label{eq:dfourier}
\tilde{\delta}_W(\vk) =\int \frac{{\rm d}^3\vp}{(2\pi)^3}\tilde{W}(\vp) \tilde{\delta}(\vk-\vp) \equiv \int_{\vp} \tilde{W}(\vp) \tilde{\delta}(\vk-\vp),
\eq
where the second equality serves to define our shorthand notation $\int_{\vp}$, which we adopt throughout. We also always assume the continuum limit in Fourier space, which is valid if the modes considered are much larger than the fundamental mode, $k \gg 2\pi/L$, where $L \sim V_W^{1/3}$ ($V_W = \int{\rm d}^3\vx W(\vx)$ is the survey volume). We further define the following estimator of the three-dimensional matter power spectrum measured in the survey
\bq\label{eq:Pk3D}
\hat{P}_W(\vk_1) = \frac{1}{V_W} \tilde{\delta}_W(\vk_1) \tilde{\delta}_W(-\vk_1).
\eq
Note that we allow this estimator to depend on the orientation of $\vk_1$, i.e., we are not restricting to the case of angle-averaged power spectra as is commonly the case in the literature. The ensemble average of this power spectrum estimator is 
\bq\label{eq:Pk3D<>}
\langle\hat{P}_W(\vk_1)\rangle &=& \frac{1}{V_W} \int_{\vp_1}\int_{\vp_2}  \tilde{W}(\vp_1)\tilde{W}(\vp_2) \langle\tilde{\delta}(\vk_1-\vp_1)\tilde{\delta}(-\vk_1-\vp_2)\rangle\nonumber \\
&=&\frac{1}{V_W}  \int_{\vp} |\tilde{W}(\vp)|^2 \Pnl(\vk_1-\vp),
\eq
where we have used the definition of the matter power spectrum $(2\pi)^3\delta_D(\vk+\vk')\Pnl(\vk) = \langle\delta(\vk)\delta(\vk')\rangle$. As mentioned above, we will always consider kinematic regimes in which the modes $\vk$ are much larger than the width of the window function in Fourier space, $k \gg 1/L$. Further, noting that for $p \gg 1/L$ the above integral is suppressed by $|W(\vp)|$, i.e.~the result is only non-negligible if $k \gg p$, we can make the approximation $\Pnl(\vk-\vp) \approx \Pnl(\vk)$ in Eq.~(\ref{eq:Pk3D<>}). Then, $\langle\hat{P}_W(\vk_1)\rangle = \Pnl(\vk_1) V_W^{-1} \int_{\vp} |\tilde{W}(\vp)|^2 = \Pnl(\vk_1)$, which demonstrates that the above estimator is unbiased for $k \gg p$.

The sample covariance of the estimator of Eq.~(\ref{eq:Pk3D<>}) can then be written as
\bq\label{eq:covdef}
\cov(\vk_1, \vk_2)&=& \langle\hat{P}_W(\vk_1)\hat{P}_W(\vk_2)\rangle - \langle\hat{P}_W(\vk_1)\rangle \langle\hat{P}_W(\vk_2)\rangle \nonumber \\
&=& \frac{1}{V_W^2}\Big[\langle\tilde{\delta}_W(\vk_1)\tilde{\delta}_W(\vk_2)\rangle\langle\tilde{\delta}_W(-\vk_1)\tilde{\delta}_W(-\vk_2)\rangle \ + \  \left(\vk_2 \leftrightarrow -\vk_2 \right)\Big] \nonumber \\
&+& \frac{1}{V_W^2}\langle\tilde{\delta}_W(\vk_1)\tilde{\delta}_W(-\vk_1)\tilde{\delta}_W(\vk_2)\tilde{\delta}_W(-\vk_2)\rangle_c.
\eq
In Appendix \ref{app:covderiv}, we analyse these correlators with some detail to help understand the impact of the survey window function in the matter covariance. Here, we display directly the final result, which can be written as
\bq\label{eq:covres}
\cov(\vk_1, \vk_2) \equiv \cov(k_1, k_2, \mu_{12}) &=& \cov^\text{G}(k_1,k_2,\mu_{12}) + \cov^\text{NG}(k_1,k_2,\mu_{12}) \nonumber \\[3pt] 
\cov^\text{G}(k_1,k_2,\mu_{12}) &=& \frac{1}{V_W^2} [\Pnl(\vk_1)]^2 \Big[ |\tilde{W}(\vk_1+\vk_2)|^2 + |\tilde{W}(\vk_1-\vk_2)|^2\Big] \nonumber \\
\cov^\text{NG}(k_1,k_2,\mu_{12}) &=& \frac{1}{V_W^2} \int_{\vp}|\tilde{W}(\vp)|^2 T_m(\vk_1, -\vk_1 + \vp, \vk_2, -\vk_2 - \vp),
\eq
where $\mu_{12} = \vk_1\vk_2/k_1/k_2$ and the trispectrum is defined as $(2\pi)^3\delta_D(\vk_{abcd})T_m(\vk_a, \vk_b, \vk_c, \vk_d) = \langle\delta(\vk_a)\delta(\vk_b)\delta(\vk_c)\delta(\vk_d)\rangle_c$. The first term on the right-hand side of Eq.~(\ref{eq:covres}) is the so-called Gaussian (G) part of the covariance, which contributes only if $|\vk_1 \pm \vk_2| \ll 1/L$; for an infinite volume, the window functions in the Gaussian term effectively work as Dirac-delta functions. Thus, the Gaussian contribution is diagonal. The second, ``NG'' term involving the trispectrum can be split into the non-Gaussian super-sample covariance (SSC) term that we wish to focus on in this paper, as well as into the connected non-Gaussian (cNG) term\footnote{Both the SSC and cNG terms are related to the connected four-point function, and hence it is somewhat inconsistent to use the word ``connected'' when describing the cNG term without the SSC contribution. We keep this terminology nonetheless, but with this small caveat in mind.}. We discuss these non-Gaussian terms in more detail next.

\subsection{The super-sample term with responses}\label{sec:ssc}

The key quantity that sets the non-Gaussian matter covariance is the trispectrum in the configuration that appears in Eq.~(\ref{eq:covres}). The mode-coupling interactions captured by this configuration of the trispectrum can be described by the following general diagram
\bq\label{eq:twgeneral}
T_m(\vk_1, -\vk_1+\vp, \vk_2, -\vk_2-\vp) \ \ =\  \ \raisebox{-0.0cm}{\includegraphicsbox[scale=1.0]{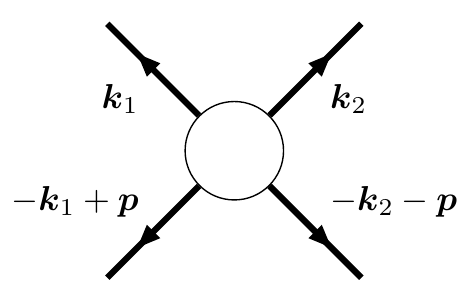}},
\eq
where the big empty blob is meant to account for all kinematically allowed vertex and loop interactions, as well as the corresponding connecting propagators, i.e.~power spectra. If $p=0$, the above trispectrum configuration reduces to the degenerate configuration of the covariance, i.e., that which corresponds to the case of infinite survey window functions. We can thus define the super sample covariance contribution as that which arises for finite small momenta $p \ll k_1, k_2$ due to \emph{large-scale correlations} of the modes with wavenumber $p$, and thus is proportional to $\Plin(p)$. That is,\footnote{Specifically, we are isolating the contribution in the $p\to 0$ limit that is non-analytic in $p$. An essentially equivalent formulation is to define the SSC term as
\bq\label{eq:Tssc_alt}
T^{\rm SSC}(\vk_1, -\vk_1, \vk_2, -\vk_2; \vp) = \lim_{p \to 0}T_m(\vk_1, -\vk_1+\vp, \vk_2, -\vk_2-\vp) \ - \ T_m^{\rm cNG}(\vk_1, -\vk_1, \vk_2, -\vk_2), 
\eq
with the limit interpreted as keeping the leading term as $p \to 0$, which are terms $\propto \Plin(p)$. This is also the formulation adopted in Sec.~II.~C of Ref.~\cite{takada/hu:2013}.}
\bq\label{eq:Tssc}
T^{\rm SSC}(\vk_1, -\vk_1, \vk_2, -\vk_2; \vp) = \left[ \lim_{p \to 0} \frac{\partial}{\partial [\Plin(p)]} T_m(\vk_1, -\vk_1+\vp, \vk_2, -\vk_2-\vp) \right] \Plin(p) .
\eq
By inspecting the possible flow of momentum inside the blob in Eq.~(\ref{eq:twgeneral}), we see that the SSC contribution has to be associated with diagrams in which there is a line (signifying a propagator $\Plin(p)$) with momentum $\vp$ connecting the {\rm left} (i.e., $\vk_1, -\vk_1+\vp$) and the {\rm right} (i.e., $\vk_2, -\vk_2-\vp$) sides; this is the only diagram proportional to $\Plin(p)$. All other diagrams are, in the limit $p\to 0$, either $(i)$ independent of $p$, and thus not of the SSC type (they belong to the connected non-Gaussian contribution); or $(ii)$ depend on higher even powers of $p$, such as $(p/k_1)^2$. Contributions of this second type are analytic in $p$ and do not involve the large-scale power spectrum $\Plin(p)$. They correspond to the smoothing of the trispectrum by the window function, and are thus not considered a part of SSC, in that they do not describe the gravitational coupling between modes of wavenumbers $\vp$ and $\vk_1, \vk_2$. Appendix \ref{app:ssctree} illustrates these arguments and provides more intuition about the derivation of the SSC term by working at tree level in perturbation theory.

Recalling that we are considering kinematic cases in which $p \lesssim 1/L \ll k_1, k_2$ (cases with $p > 1/L$ will be suppressed by the window function and cases with $k_1, k_2 \sim 1/L$ are dominated by the Gaussian contribution), then the diagrams contributing to the SSC term defined above correspond to the interaction of a soft mode $\vp$ with two hard modes $\vk_1, \vk_2$, which are two interactions of the $\R_1$ type. Within the response approach, we can thus write the SSC term as a single diagram:
\bq\label{eq:sscdiagrams}
T^{\rm SSC}(\vk_1, -\vk_1, \vk_2, -\vk_2; \vp) &=& \raisebox{-0.0cm}{\includegraphicsbox[scale=0.85]{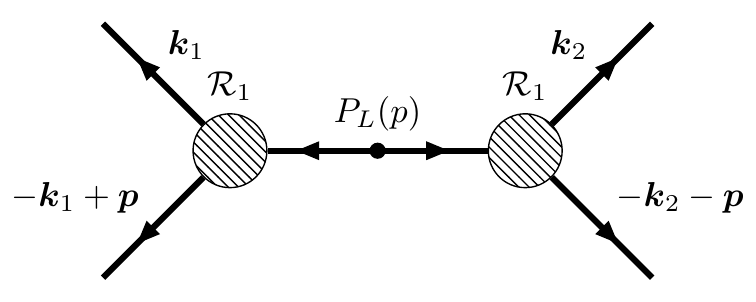}} \nonumber \\
&=& \R_1(k_1, -\mu_{\vp,\vk_1}) \R_1(k_2, \mu_{\vp,\vk_2})\Pnl(k_1)\Pnl(k_2)\Plin(p).
\eq
In this diagram, one of the vertices has incoming $\vp$ momentum while the other has $-\vp$, hence the different signs in the second argument of the two $\R_1$ functions; numerically, this is however irrelevant because $\R_1$ is proportional to the square of this cosine angle (cf.~Eq.~(\ref{eq:R1dec})).

Returning to the calculation of the covariance, the non-Gaussian part in Eq.~(\ref{eq:covres}) can then be written as
\bq\label{eq:covNG}
\cov^{\rm NG}(k_1, k_2, \mu_{12}) &=& \frac{1}{V_W}T_m^{\rm cNG}(\vk_1, -\vk_1, \vk_2, -\vk_2) + \frac{1}{V_W^2}\int_{\vp}|\tilde{W}(\vp)|^2T_m^{\rm SSC}(\vk_1, -\vk_1, \vk_2, -\vk_2; \vp) \nonumber \\
&=& \cov^{\rm cNG} (k_1, k_2, \mu_{12}) + \cov^{\rm SSC} (k_1, k_2, \mu_{12}),
\eq
where the second line establishes our notation to denote the cNG and SSC matter covariance terms, and we have neglected the smoothing effect of the window function on the cNG term. Using Eqs.~(\ref{eq:R1dec}) and (\ref{eq:sscdiagrams}), the SSC term can be written explicitly as
\bq\label{eq:cov_ssc}
\cov^{\rm SSC}(k_1, k_2, \mu_{12})&=& \cov^{\rm SSC}_{\ssca} + \cov^{\rm SSC}_{\sscb} + \cov^{\rm SSC}_{\sscc} + \cov^{\rm SSC}_{\sscd},
\eq
with
\bq 
\label{eq:covsscA} V_W^2 \frac{\cov^{\rm SSC}_{\ssca}(\vk_1,\vk_2)}{P_m(k_1)P_m(k_2)} &=& R_1(k_1)R_1(k_2) \int_{\vp}  |\tilde{W}(\vp)|^2 \Plin(p) \\
\label{eq:covsscB} V_W^2 \frac{\cov^{\rm SSC}_{\sscb}(\vk_1,\vk_2)}{P_m(k_1)P_m(k_2)} &=& R_K(k_1)R_K(k_2) \int_{\vp}  |\tilde{W}(\vp)|^2 \left(\mu_{\vp,\vk_1}^2 - \frac{1}{3}\right) \left(\mu_{\vp,\vk_2}^2 - \frac{1}{3}\right) \Plin(p) \\
\label{eq:covsscC} V_W^2 \frac{\cov^{\rm SSC}_{\sscc}(\vk_1,\vk_2)}{P_m(k_1)P_m(k_2)} &=& R_1(k_1)R_K(k_2) \int_{\vp}  |\tilde{W}(\vp)|^2  \left(\mu_{\vp,\vk_2}^2 - \frac{1}{3}\right) \Plin(p) \\
\label{eq:covsscD} V_W^2 \frac{\cov^{\rm SSC}_{\sscd}(\vk_1,\vk_2)}{P_m(k_1)P_m(k_2)} &=& R_K(k_1)R_1(k_2) \int_{\vp}  |\tilde{W}(\vp)|^2 \left(\mu_{\vp,\vk_1}^2 - \frac{1}{3}\right) \Plin(p)\,,
\eq
where $\mu_{\vp,\vk_i}= \vp\vk_i/(p k_i)$, $i \in \{1,2\}$. The standard SSC term derived in Ref.~\cite{takada/hu:2013} corresponds to the $\cov^{\rm SSC}_{\ssca}$ part above, which describes the covariance due to the presence of a super-survey density perturbation. The other terms involve $R_K$ and therefore correspond to the covariance that is induced by super-survey tidal fields.  The covariance of the angle-averaged (monopole) three-dimensional matter power spectrum,
  \be\label{eq:eqmonoPw}
\hat P_W(k) \equiv \int \frac{d^2 \hat{\vk}}{4\pi} \hat P_W(\vk)
  \ee
is obtained from the above expressions by averaging over the directions of $\vk_1$ and $\vk_2$. In this case, the contributions involving the tidal response $R_K$ vanish exactly, \emph{irrespective of the shape and size of the survey window function.} For higher power spectrum multipoles (e.g.~quadrupole terms that arise via RSD in the case of galaxies as tracers), the tidal response contributions do not vanish and can induce additional anisotropies as discussed in Refs.~\cite{akitsu/takada/li, 2017arXiv171100012A, 2017arXiv171100018L}.

\section{Lensing covariance}\label{sec:lensing}

The expressions derived in the previous section are for the three-dimensional matter power spectrum, which is not directly observable. Current and future large-scale structure surveys instead measure biased and redshift space distorted (RSD) versions of the matter density in the case of galaxy clustering, and the projected matter density field in the case of weak gravitational lensing. In this section, we focus on the latter.

We begin with a derivation that assumes the flat-sky and Limber approximations for all modes involved, and whose derivation steps are analogous to those taken in the previous section. We then derive SSC formulae that are valid for curved-sky and that go beyond the Limber approximation in the super-survey density and tidal field perturbations. In the section after this one, we will compare the results from these two derivations.

\subsection{Lensing SSC in flat sky and with Limber's approximation}\label{sec:lenscovderiv_1}

In General Relativity, the lensing convergence $\kappa$ (see Refs.~\cite{2001PhR...340..291B, 2005astro.ph..9252S, 2008ARNPS..58...99H, 2015RPPh...78h6901K} for weak gravitational lensing reviews) can be given in terms of a weighted projection of the three dimensional density field as
\bq\label{eq:kappadef}
\kappa(\vtheta) = \int {\rm d}\chi\  g(\chi)\  \delta(\vx = \chi\vtheta, z(\chi)),
\eq
where $\vtheta$ is a two-dimensional position vector defined on the plane of the sky; from hereon we will write $z(\chi)$ simply as $z$ to ease the notation. Here, we have neglected higher-order corrections such as beyond-Born \cite{2017PhRvD..95l3503P, 2010PhRvD..81h3002B}, reduced-shear \cite{dodelson/shapiro/white,krause/hirata:2010} and lensing-bias contributions \cite{schmidt/etal:2009}. We have also assumed a spatially flat universe and replaced the two-dimensional Laplacian defined on the sky by the three-dimensional one to relate the lensing convergence to the three-dimensional density contrast using the Poisson equation; this is however a valid approximation in the small-scale, flat-sky limit assumed here. The variable $\chi \equiv \chi(z)$ is the comoving distance out to redshift $z$ ($\chi = c \int_0^z {\rm d}z'/(H(z')$ in a spatially flat universe) and the function $g(\chi)$ is often called the lensing kernel or efficiency: for a single lensing source plane at $\chi_S$ it is given by $g(\chi, \chi_S) = (3H_0^2\Omega_m/2/c^2) (1+z) (\chi_S-\chi)\chi/\chi_S$. 

Analogously to Eq.~(\ref{eq:dreal}), we can define the observed convergence field as
\bq\label{eq:kappaobs}
\kappa_\W(\vtheta) = \W(\vtheta) \kappa(\vtheta),
\eq
where $\W(\vtheta)$ is now a window function defined on the sky that is unity within the surveyed area (solid angle $\Omega_W$) and zero outside. Under the flat-sky approximation, the Fourier transform of the observed convergence reads (defining also a short-hand notation for two-dimensional volume integrals in Fourier space)
\bq\label{eq:dfourierproj}
\tilde{\kappa}_\W(\vell) = \int\frac{{\rm d}^2\vell'}{(2\pi)^2}\tilde\W(\vell')\tilde{\kappa}(\vell - \vell') \equiv \int_{\vell'}\tilde\W(\vell')\tilde{\kappa}(\vell - \vell'),
\eq
with a power spectrum estimator that can be written as
\bq\label{eq:Clestimator}
\hat{C}(\vell) = \frac{1}{\Omega_\W} \tilde{\kappa}_\W(\vell)\tilde{\kappa}_\W(-\vell).
\eq
Following the same reasoning as in after Eq.~(\ref{eq:Pk3D<>}), this estimator is indeed an unbiased measure of the true power spectrum $C(\vell)$ defined as $(2\pi)^2\delta_D(\vell+\vell')C(\vell) = \langle\tilde{\kappa}(\vell)\tilde{\kappa}(\vell')\rangle$ for $\ell \gg 2\pi/\theta_0$, where $\theta_0 \sim \Omega_\W^{1/2}$ is the typical survey window size.

The sample lensing covariance can then be written as
\bq\label{eq:lenscovdef}
\covkappa(\vell_1, \vell_2) &=& \langle\hat{C}(\vell_1)\hat{C}(\vell_2)\rangle - \langle\hat{C}(\vell_1)\rangle\langle\hat{C}(\vell_2)\rangle \nonumber \\
&=& \frac{1}{\Omega_\W^2} \Big[\langle\tilde{\kappa}_\W(\vell_1)\tilde{\kappa}_\W(\vell_2)\rangle\tilde{\kappa}_\W(-\vell_1)\tilde{\kappa}_\W(-\vell_2)\rangle \ + \left(\vell_2 \leftrightarrow -\vell_2\right)\Big] \nonumber \\
&+& \frac{1}{\Omega_\W^2} \langle\tilde{\kappa}_\W(\vell_1)\tilde{\kappa}_\W(-\vell_1)\tilde{\kappa}_\W(\vell_2)\tilde{\kappa}_\W(-\vell_2)\rangle_c.
\eq
Following similar steps as for the three-dimensional case, we straightforwardly arrive at
\bq
\covkappa(\vell_1, \vell_2) &=& \frac{1}{\Omega_\W^2}[C(\vell_1)]^2 \Big[|\tilde\W(\vell_1+\vell_2)|^2 + |\tilde\W(\vell_1-\vell_2)|^2\Big] \nonumber \\
&+& \frac{1}{\Omega_\W^2} \int_{\vell} |\tilde\W(\vell)|^2 \T_{\kappa}(\vell_1, -\vell_1 + \vell, \vell_2, -\vell_2 - \vell),
\eq
with $(2\pi)^2\delta_D(\vell_{abcd})\T_\kappa(\vell_a,\vell_b,\vell_c,\vell_d) = \langle\tilde{\kappa}(\vell_a)\tilde{\kappa}(\vell_b)\tilde{\kappa}(\vell_c)\tilde{\kappa}(\vell_d)\rangle_c$ defining the trispectrum of the convergence field. We refer to the first and second terms on the right-hand side as the Gaussian, $\covkappa^{\rm G}$, and non-Gaussian contributions to the convergence covariance, respectively.  Under the Limber approximation, the convergence power spectrum and trispectrum can be related to those of the three-dimensional density contrast as (cf.~Appendix \ref{app:limber} for the derivation)
\bq
\label{eq:limberapprox_C} C(\vell) &=& \int{\rm d}\chi [g(\chi)]^2 \chi^{-2} \Pnl\left(\vk_{\ell}, z\right), \\
\label{eq:limberapprox_T} \T_\kappa(\vell_a,\vell_b,\vell_c,\vell_d) &=& \int{\rm d}\chi [g(\chi)]^4 \chi^{-6} T_m\left(\vk_{\ell_a}, \vk_{\ell_b}, \vk_{\ell_c}, \vk_{\ell_d}, z\right)\,,
\eq
where here and throughout $\vk_{\ell} \equiv (\vell+1/2)/\chi$. Using the split of the matter trispectrum into its degenerate and super-sample parts, the Gaussian, cNG and SSC contributions to the lensing convergence covariance can be given, respectively, as
\bq\label{eq:lenscovbits}
\label{eq:covkappa_G}\covkappa^{\rm G}(\vell_1, \vell_2) &=& \frac{1}{\Omega_\W^2}[C(\vell_1)]^2 \Big[|\tilde\W(\vell_1+\vell_2)|^2 + |\tilde\W(\vell_1-\vell_2)|^2\Big], \\[3pt]
\label{eq:covkappa_NG}\covkappa^{\rm cNG}(\vell_1, \vell_2) &=& \frac{1}{\Omega_\W}\int{\rm d}\chi [g(\chi)]^4 \chi^{-6} T_m^{\rm cNG}\left(\vk_{\ell_1}, -\vk_{\ell_1}, \vk_{\ell_2}, -\vk_{\ell_2}; z\right), \\[3pt]
\label{eq:covkappa_SSC}\covkappa^{\rm SSC}(\vell_1, \vell_2) &=& \frac{1}{\Omega_\W^2}\int{\rm d}\chi [g(\chi)]^4 \chi^{-6} \int_{\vell} |\tilde{\W}(\vell)|^2 T^{\rm SSC}\left(\vk_{\ell_1}, -\vk_{\ell_1}, \vk_{\ell_2}, -\vk_{\ell_2}, \vk_{\ell}\right), \nonumber \\[3pt]
&=& \covkappa^{\rm SSC}_{\ssca} + \covkappa^{\rm SSC}_{\sscb} + \covkappa^{\rm SSC}_{\sscc} + \covkappa^{\rm SSC}_{\sscd}.
\eq
In Eq.~(\ref{eq:covkappa_SSC}), we have again split the SSC contribution into four pieces given by
\bq\label{eq:SSCpieces}
\label{eq:kappassc_A} \covkappa^{\rm SSC}_{\ssca}(\vell_1,\vell_2) &=& \frac{1}{\Omega_\W^2} \int {\rm d}\chi [g(\chi)]^4\chi^{-6} R_1\left(k_{\ell_1}, z\right)R_1\left(k_{\ell_2}, z\right) \Pnl\left(k_{\ell_1}, z\right)\Pnl\left(k_{\ell_2}, z\right) \nonumber \\
&&   \times \int_{\vell} |\tilde{\W}(\vell)|^2 \Plin\left(k_{\ell}, z\right), \\
\label{eq:kappassc_B} \covkappa^{\rm SSC}_{\sscb}(\vell_1,\vell_2) &=& \frac{1}{\Omega_\W^2} \int {\rm d}\chi [g(\chi)]^4\chi^{-6} R_K\left(k_{\ell_1}, z\right)R_K\left(k_{\ell_2}, z\right) \Pnl\left(k_{\ell_1}, z\right)\Pnl\left(k_{\ell_2}, z\right) \nonumber \\
&&   \times \int_{\vell} |\tilde{\W}(\vell)|^2 \left(\mu_{\vell,\vell_1}^2 - \frac{1}{3}\right) \left(\mu_{\vell,\vell_2}^2 - \frac{1}{3}\right) \Plin\left(k_{\ell}, z\right), \\
\label{eq:kappassc_C} \covkappa^{\rm SSC}_{\sscc}(\vell_1,\vell_2) &=& \frac{1}{\Omega_\W^2} \int {\rm d}\chi [g(\chi)]^4\chi^{-6} R_1\left(k_{\ell_1}, z\right)R_K\left(k_{\ell_2}, z\right) \Pnl\left(k_{\ell_1}, z\right)\Pnl\left(k_{\ell_2}, z\right) \nonumber \\
&&   \times \int_{\vell} |\tilde{\W}(\vell)|^2 \left(\mu_{\vell,\vell_2}^2 - \frac{1}{3}\right) \Plin\left(k_{\ell}, z\right), \\
\label{eq:kappassc_D} \covkappa^{\rm SSC}_{\sscd}(\vell_1,\vell_2) &=& \frac{1}{\Omega_\W^2} \int {\rm d}\chi [g(\chi)]^4\chi^{-6} R_K\left(k_{\ell_1}, z\right)R_1\left(k_{\ell_2}, z\right) \Pnl\left(k_{\ell_1}, z\right)\Pnl\left(k_{\ell_2}, z\right) \nonumber \\
&&   \times \int_{\vell} |\tilde{\W}(\vell)|^2 \left(\mu_{\vell,\vell_1}^2 - \frac{1}{3}\right) \Plin\left(k_{\ell}, z\right),
\eq
where $\mu_{\vell,\vell_1} = \vell\vell_1/\ell\ell_1$, $\mu_{\vell\vell_2} = \vell\vell_2/\ell\ell_2$ and $z \equiv z(\chi)$. Recall that $\vell, \vell_i$ are 2D vectors defined on the flat-sky plane. The monopole of the convergence power spectrum is defined as
\bq\label{eq:Clestimatoraa1}
\hat{C}(\ell_1) = \int \frac{d\varphi_{\vell_1}}{2\pi} \hat{C}(\vell_1)\,.
\eq
The SSC contribution to the covariance of this monopole power spectrum is obtained by angle-averaging Eq.~(\ref{eq:covkappa_SSC}) as 
\bq\label{eq:covkappa_SSC_mono_flat}
\covkappa^{\rm SSC}(\ell_1, \ell_2) &=& \int_{0}^{2\pi}\int_{0}^{2\pi}\frac{{\rm d}\varphi_{\vell_1}{\rm d}\varphi_{\vell_2}}{(2\pi)^2} \covkappa^{\rm SSC}(\vell_1, \vell_2) \\
&=& \frac{1}{\Omega_\W^2} \int_{\vell} |\tilde{\W}(\vell)|^2 \sigma_{\ell_1, \ell_2}^{\ell, {\rm flat}},
\eq
with
\bq\label{eq:sigma_flat}
\sigma_{\ell_1, \ell_2}^{\ell, {\rm flat}} &=& \int {\rm d}\chi \frac{[g(\chi)]^4}{\chi^6} \left(R_1\left(k_{\ell_1}, z\right) + \frac{R_K\left(k_{\ell_1}, z\right)}{6} \right)  \left(R_1\left(k_{\ell_2}, z\right) + \frac{R_K\left(k_{\ell_2}, z\right)}{6} \right) \nonumber \\
&& \times \Pnl\left(k_{\ell_1}, z\right)\Pnl\left(k_{\ell_2}, z\right) \Plin\left(k_{\ell}, z\right),
\eq
where $\varphi_{\vell_i}$ is the polar angle of the two-dimensional vector $\vell_i$. An interesting point to note from the above equation is that, contrary to the three-dimensional case, the tidal response function $R_K$ contributes to the lensing SSC even after a full angle-average. Numerically, this can be traced back to the fact that the angle averages being performed are now two-dimensional, under which the $\mu^2-1/3$ terms do not vanish (as they do in three-dimensional angle averages). Physically, as we will see with more detail in the next subsection, this follows from the fact that a super-survey tidal field can contribute monopole terms via the trace of its two-dimensional projection onto the sky, as well as its projection along the line-of-sight.

\subsection{Lensing SSC in curved sky and beyond Limber's approximation in super-survey modes}\label{sec:lenscovderiv_2}

The expressions derived in the previous subsection assume flat sky, as well as the Limber approximation for all modes involved. This is a potential concern for the accuracy of the SSC contribution, since the effect of super-survey modes involved is not necessarily adequately described by the Limber and flat-sky limits, for a survey covering a significant fraction of the sky. In this subsection, we present an alternative derivation of the lensing SSC contribution that is valid in curved-sky and goes beyond the Limber approximation in the super-survey mode. 

As in the previous subsection, consider a lensing survey with footprint $\W(\vtheta)$ that extends out to a single source redshift $z_S$. Additionally, let us consider that this lensing lightcone is embedded in a region of the Universe with long-wavelength density and tidal fluctuations. To linear order, the latter are completely characterized by the tensor
\bq\label{eq:Pi_ij}
\Pi_{ij}(\vx, z) &=& K_{ij}(\vx, z) + \frac{\delta_{ij}}{3}\delta(\vx, z) \nonumber \\
 &=& D(z) \left[K_{ij}(\vx) + \frac{\delta_{ij}}{3}\delta(\vx) \right],
\eq
where $\delta(\vx, z)$ and $K_{ij}(\vx, z)$ describe the super-survey density and tidal field. The second line factors out the linear growth factor (normalized as $D(z=0) = 1$); if $z$ is omitted in the arguments, then $z=0$ is implicitly assumed for ease of notation. The quantity $\Pi_{ij}$ can be projected separately along the line-of-sight direction $\hat{\vn}$, as well as on the sky, respectively, as 
\bq\label{eq:Pi_ij_projections}
\Pi_{\parallel}(\vx) &=& \hat{\vn}^i\hat{\vn}^j \Pi_{ij}(\vx), \nonumber \\
\Pi_{\perp, ij}(\vx) &=& \P^{k}_{i}\P^{l}_{j} \Pi_{kl}(\vx),
\eq
where $\P_{ij} = \delta_{ij} - \hat{\vn}_i\hat{\vn}_j$ is the projection operator onto the sky and $\hat{\vn} \equiv \hat{\vn}(\vtheta)$ is a three-dimensional unit vector at the observer that points at the sky coordinate $\vtheta$. The tensor projection on the sphere can be further decomposed into a trace and a traceless part, respectively, as
\bq\label{eq:Pi_ij_perp_traces}
\Pi_{\perp}^{\rm trace}(\vx) &=& \P^{ij} \Pi_{\perp, ij}(\vx), \nonumber \\ 
\Pi_{\perp, ij}^{\rm traceless}(\vx) &=& \Pi_{\perp, ij}(\vx) - \frac12 \P_{ij}\Pi_{\perp}^{\rm trace}(\vx).
\eq
Equations (\ref{eq:Pi_ij_projections}) and (\ref{eq:Pi_ij_perp_traces}) will be used below to characterize separate contributions from the long-wavelength super-survey mode.

We now wish to determine how the convergence power spectrum measured in some sub-patch of the footprint gets modified by the presence of the long-wavelength perturbation. The shape and exact size $\Delta\theta$ of these sub-patches is not important (we will later average over the whole survey footprint $\W(\vtheta)$). The only constraint imposed on the patch size $\Delta\theta$ is that it is larger than the angular scales on which the lensing power spectrum is measured, i.e. $\Delta\theta > 1/\ell$. We will also assume that $\ell$ is sufficiently large so that we can use the Limber approximation for the power spectrum (cf.~Eq.~(\ref{eq:limberapprox_C})), i.e. $\ell \gtrsim 20$ \cite{2007AA...473..711S, 2008PhRvD..78l3506L,threepointlens,2017JCAP...05..014L}. Equation~(\ref{eq:limberapprox_C}) essentially states that the lensing convergence power spectrum is obtained by superimposing the three-dimensional matter power spectrum along the line-of-sight with a geometrical weight given by the lensing kernel. We thus evaluate the convergence power spectrum in the presence of the long-wavelength perturbation using Eq.~(\ref{eq:limberapprox_C}), but replacing $P_m(k, z)$ by the corresponding quantity modulated by super-survey density and tidal fields:
\bq\label{eq:modPm}
P_m(\vk, z | \Pi(\vx)) = P_m(k,z) \left[1 + R_{1}(k, z)\delta(\vx,z) + R_{K}(k, z)\hat{\vk}^{i}\hat{\vk}^{j} K_{ij}(\vx, z)\right].
\eq
The left-hand side of this equation should be understood as the local 3D matter power spectrum measured in a patch of size $\lambda > 1/k$ centered around the position $\vx$. The right-hand side assumes that the modes $\vp$ contributing to $\d(\vx),\,K_{ij}(\vx)$ are of much larger scale than the patch size, $1/p \gg \lambda$. We can then write
\bq\label{eq:Cl_pi_1}
C(\vell, \hat{\vn} | {\Pi}) - C(\vell) &=& \int_{0}^{\chi_S} {\rm d}\chi [g(\chi)]^2 \chi^{-2} \Pnl\left(k_{\ell}, z\right) \nonumber \\ 
&& \times D(z) \int_{\vp} \left[R_{1}(k_{\ell}, z)\delta(\vp) + R_{K}(k_{\ell}, z)\hat{\vk}_{\ell}^{i}\hat{\vk}_{\ell}^{j} K_{ij}(\vp)\right] e^{i\vp\vx},
\eq
where $C(\vell, \hat{\vn} | {\Pi})$ is the convergence power spectrum in a small patch around the direction $\hat{\vn}$ in the presence of super-survey fluctuations, and $\vx = \chi \hat{\vn}$. We have also expanded the long-wavelength density and tidal fields at $\vx$ in Fourier modes. Note also that we adopt the Limber approximation for sub-survey modes contributing to the angular power spectrum, and correspondingly evaluate the power spectrum responses at $k = k_{\ell} \equiv (\ell+1/2)/\chi$. 

To continue our derivation, we make the assumption that the wavelength of the super-survey mode is sufficiently large that it can be approximated as constant across the small patch where we are measuring the power spectrum. This is in keeping with \refeq{modPm}, since the angular size $\Delta\theta > 1/\ell$ corresponds to a transverse patch size at comoving distance $\chi$ of $\lambda_\perp = \chi\Delta\theta > \chi/\ell \approx 1/k_\ell$. Correspondingly, $p \ll 1/\lambda_\perp$ implies that $p \ll 1/(\chi\Delta\theta)$, which is the transverse physical length scale of the patch on the sky. This allows us to write $e^{i\vp\vx} \approx e^{i\hat{\vp}\hat{\vn}p\chi}$. Note that we do not assume that the perturbation is constant along the line of sight, over which we integrate. Further noting that $K_{ij}(\vp) = (\hat{\vp_i}\hat{\vp_j} - \delta_{ij}/3)\delta(\vp)$, we have $\delta = \Pi_{\parallel} + \Pi_{\perp}^{\rm trace}$ and $\hat{\vk}_{\ell}^{i}\hat{\vk}_{\ell}^{j} K_{ij} = \Pi_{\perp}^{\rm trace}/6 - \Pi_{\parallel}/3 + \hat{\vk}_{\ell}^{i}\hat{\vk}_{\ell}^{j}\Pi_{\perp, ij}^{\rm traceless}$, with which we can write

\bq\label{eq:Cl_pi_2}
C(\vell, \hat{\vn} | {\Pi})  &=& C^{(0)}(\ell, \hat{\vn}  | {\Pi}) + C^{(2)}(\vell, \hat{\vn}  | {\Pi})
\eq
with
\bq\label{eq:Cl_pi_mono}
C^{(0)}(\ell, \hat{\vn}  | {\Pi}) - C(\ell) &=& \int_{0}^{\chi_S} {\rm d}\chi [g(\chi)]^2 \chi^{-2} D(z) \Pnl\left(k_{\ell}, z\right), \nonumber \\ 
&& \times \int_{\vp} \left[ R_{\perp}(k_{\ell}, z) \Pi_{\perp}^{\rm trace}(\vp) +R_{\parallel}(k_{\ell}, z) \Pi_{\parallel}(\vp)  \right] e^{i\hat{\vp}\hat{\vn}p\chi}
\eq
\bq\label{eq:Cl_pi_quad}
C^{(2)}(\vell, \hat{\vn}  | {\Pi}) &=& \int_{0}^{\chi_S} {\rm d}\chi [g(\chi)]^2 \chi^{-2} D(z) \Pnl\left(k_{\ell}, z\right) \int_{\vp}  R_{K}(k_{\ell}, z) \hat{\vk}_{\ell}^{i}\hat{\vk}_{\ell}^{j}\Pi_{\perp, ij}^{\rm traceless} e^{i\hat{\vp}\hat{\vn}p\chi}, \nonumber \\
\eq
and where 
\bq
\quad R_{\perp} = R_1 + R_K/6 \quad \mbox{,}\quad R_{\parallel} = R_1 - R_K/3\,.\nonumber
\eq
The terms $C^{(0)}(\ell, \hat{\vn}  | {\Pi})$ and $C^{(2)}(\vell, \hat{\vn}  | {\Pi})$ represent a monopole and a quadrupole contribution, respectively, that are induced by the long-wavelength perturbation $\Pi(\vp)$. Since $C^{(2)}$ vanishes after angle-averaging over $\varphi_{\vell}$, regardless of the shape of the survey window function, we will focus on $C^{(0)}(\vell | {\Pi})$ in the following. Note, however, that Eq.~(\ref{eq:Cl_pi_quad}) suggests that a measurement of the \emph{anisotropic} part of the lensing power spectrum on the sky (e.g. in sub-patches of the survey) will allow for a measurement of the sky-projected tidal field, in analogy with similar approaches applied to galaxy power spectra in \cite{pen/etal:2012,zhu/etal:2016}. 

Noting that
\bq\label{eq:auxeqs}
\Pi_{\perp}^{\rm trace}(\vp) e^{i\hat{\vp}\hat{\vn}p\chi} &=& \left[1-(\hat{\vp}\hat{\vn})^2\right]\delta(\vp)e^{i\hat{\vp}\hat{\vn}p\chi} = \delta(\vp) \left[1+\partial^2_x\right] e^{i\hat{\vp}\hat{\vn}x}, \\
\Pi_{\parallel}(\vp) e^{i\hat{\vp}\hat{\vn}p\chi} &=& (\hat{\vp}\hat{\vn})^2\delta(\vp)e^{i\hat{\vp}\hat{\vn}p\chi} = - \delta(\vp) \partial^2_x e^{i\hat{\vp}\hat{\vn}x}, \\
e^{i\hat{\vp}\hat{\vn}x} &=& 4\pi \sum_{LM} i^L j_{L}(x) Y_{LM}^*(\hat{\vp}) Y_{LM}(\hat{\vn}),
\eq
we can expand the monopole term in spherical harmonics as
\bq\label{eq:Cl_pi_mono_2}
C^{(0)}(\ell, \hat{\vn}  | {\Pi}) - C(\vell) &=& \sum_{LM} a_{LM}(\ell) Y_{LM}(\hat{\vn}) \\
a_{LM}(\ell) &=& 4\pi i^L \int_{\vp} \delta(\vp) Y_{LM}^*(\hat{\vp}) f_\ell^L(p) \\
\label{eq:flLp}f_\ell^L(p) &=&  \int_{0}^{\chi_S} {\rm d}\chi [g(\chi)]^2 \chi^{-2} D(z) \Pnl\left(k_{\ell}, z\right) \left[R_{\perp}(k_{\ell}, z) + \frac12 R_{K}(k_{\ell}, z) \partial^2_x\right] j_L(x) \nonumber \\
\eq
where $x = p\chi$, $\partial^2_x = \partial^2/\partial x^2$ and $^*$ denotes complex conjugation. The terms $\propto \partial^2_x$ account for the fact that, when going beyond Limber's approximation, the super-survey mode $\vp$ is not necessarily perpendicular to the line-of-sight direction, i.e., $\hat{\vn}\hat{\vp} \neq 0$. The angle-averaged convergence power spectrum in the presence of a long-wavelength mode averaged over the whole survey footprint can then be written as
\bq\label{eq:Cl_pi_mono_3}
C^{(0)}(\ell| {\Pi}) &=& \frac{1}{\Omega_\W} \int {\rm d}^2\hat{\vn} \W(\hat{\vn}) C^{(0)}(\ell, \hat{\vn} | {\Pi}) \nonumber \\
&=& C(l) + \frac{1}{\Omega_\W} \sum_{LM} b_{LM} a_{LM}(\ell),
\eq
where $b_{LM}$ are the spherical harmonic coefficients of the window function, i.e., $\W(\hat{\vn}) = \sum_{LM}b_{LM} Y_{LM}(\hat{\vn})$ and we have used the orthogonality of the spherical harmonics in the second equality.

Finally, the SSC contribution can be obtained by taking the variance of the second term on the right-hand side of Eq.~(\ref{eq:Cl_pi_mono_3}), which captures the leading effect of survey-scale modes (note that $\langle a_{LM}(p) \rangle = 0$)
\bq\label{eq:ssc_bL_1}
\cov^{\rm SSC}(\ell_1, \ell_2) &=& \frac{1}{\Omega_\W^2} \sum_{LM}\sum_{L'M'} b_{LM}b^*_{L'M'} \langle a_{LM}(\ell_1)a^*_{L'M'}(\ell_2)\rangle \nonumber \\
&=& \frac{1}{\Omega_\W^2} \sum_{LM}\sum_{L'M'} b_{LM}b^*_{L'M'} \int_{\vp}\int_{\vp'}\langle\delta(\vp)\delta(-\vp')\rangle f_{\ell_1}^L(p)f_{\ell_2}^L(p') Y_{LM}(\hat{\vp})Y^*_{L'M'}(\hat{\vp'}) \nonumber \\
&=& \frac{1}{\Omega_\W^2} \sum_{LM}\sum_{L'M'} b_{LM}b^*_{L'M'} \frac{2}{\pi} \int {\rm d}pp^2 \Plin(p)f_{\ell_1}^L(p)f_{\ell_2}^L(p) \delta_{LL'}\delta_{MM'} \nonumber \\
&=& \frac{1}{\Omega_\W^2} \sum_{LM} |b_{LM}|^2 \sigma_{\ell_1,\ell_2}^L
\eq
where
\bq\label{eq:sigmal1l2L}
\sigma_{\ell_1,\ell_2}^L = \frac{2}{\pi} \int {\rm d}p\:p^2 \Plin(p)f_{\ell_1}^L(p)f_{\ell_2}^L(p).
\eq
In Eq.~(\ref{eq:ssc_bL_1}) we have used again the orthogonality of the spherical harmonics, as well as $\langle\delta(\vp)\delta(-\vp')\rangle = (2\pi)^3\delta_D(\vp + \vp') P_{\rm L}(p)$; the subscript ${\rm _L}$ in the linear power spectrum should not be confused with the index $L$ and, recall, $\Plin(p) \equiv \Plin(p, z = 0)$. Using the above equations, one can confirm that for $L=0$, corresponding to an all-sky survey, the integrand of Eq.~(\ref{eq:sigmal1l2L}) is suppressed when $p$ becomes larger than a few times $1/\chi_S$, as it should be if $\vp$ is meant to describe a super-survey density/tidal fluctuation.

Equations (\ref{eq:ssc_bL_1}) and (\ref{eq:sigmal1l2L}) represent our main result in this section, which corresponds to a calculation of the lensing SSC contribution that is valid for any survey footprint on the (curved) sky and that goes beyond the Limber approximation in the long wavelength super-survey modes. Equations (\ref{eq:ssc_bL_1}) and (\ref{eq:sigmal1l2L}) show that tidal fields contribute to the monopole lensing SSC both through $R_\perp$ and $R_\parallel$. This can be traced back to the fact that $K_{ij}(\vp)$ contributes both to $\Pi_\perp^{\rm trace}$ and $\Pi_\parallel$ in Eq.~(\ref{eq:Cl_pi_mono}).

In fact, one can show analytically that our curved-sky expression reduces to the flat-sky result in the limit of small windows on the sky, where the sum over $L,M$ in \refeq{ssc_bL_1} is dominated by high $L$ (cf.~the lower right panel of Fig.~\ref{fig:fsky1}). In this limit, only the $R_\perp$ contribution is relevant. For sufficiently large $L$, Eq.~(\ref{eq:sigmal1l2L}) can be schematically written as (dropping the contribution from terms involving $\partial_x^2$, which are suppressed in this limit)
\bq
\sigma_{\ell_1\ell_2}^L \sim \int{\rm d\chi}\int{\rm d\chi'}\int{\rm d}pp^2 j_{L}(p\chi)j_{L}(p\chi') \left[\cdots\right]  \stackrel{L\gg1}{\approx}  \int{\rm d\chi}\int{\rm d\chi'} \frac{\pi}{2\chi^2}\delta_D\left({\chi - \chi'}\right) \left[\cdots\right],
\nonumber
\eq
where the approximation holds if the spherical Bessel functions are rapidly oscillating compared to the rest of the integrand, represented by the $\left[\cdots\right]$. Carrying out these steps, together with the approximation that for high $L$, $\int_{\vell} \left| \mathcal{W}(\vell) \right|^2 \approx \sum_{LM}\left|b_{LM}\right|^2$, one can show that the curved-sky result approaches the flat-sky expression as $f_{\rm sky} \to 0$, i.e. in the small-survey limit, as expected. In the next section, we will demonstrate this explicitly via numerical evaluation. 

One should note that the derivation in this subsection does assume the Limber approximation for the modes contributing to the angular power spectrum $C(\ell_i)$. Hence, the result is not guaranteed to be a good approximation on multipoles $\ell_1, \ell_2 \lesssim 20$. As we will see below, however, on such large angular scales, the signal-to-noise ratio is completely dominated by the Gaussian contribution, making any inadequacy in the SSC description for $\ell_1, \ell_2 \lesssim 20$ irrelevant in practice. One should also note that Ref.~\cite{2016arXiv161205958L} presented SSC formulae similar to Eqs.~(\ref{eq:ssc_bL_1}) and (\ref{eq:sigmal1l2L}), but without the contribution from the $R_K$ and $\sim \partial_x^2 j_L(x)$ terms in Eq.~(\ref{eq:flLp}).

\section{Quantitative results}\label{sec:results}

In this section, we present numerical results from the equations derived in the previous section for lensing. We begin with a comparison between the two SSC derivations of Secs.~\ref{sec:lenscovderiv_1} and \ref{sec:lenscovderiv_2}. Then,  we compare the size of the SSC contribution against those of the G and cNG terms. We adopt a $\lcdm$ cosmology with parameters $\Omega_mh^2 = 0.14695$, $\Omega_bh^2 = 0.02205$, $h = 0.70$, $n_s=0.96$, $\sigma_8(z=0) = 0.83$.  We evaluate linear and nonlinear three-dimensional matter power spectra using {\sc CAMB} \cite{camb} and the {\sc COYOTE} emulator \cite{emulator}, respectively. We show only the monopole part of the lensing covariance for a single source redshift of $z_S = 1$.

We consider the angle-averaged power spectrum estimator in angular wavenumber bins $\Delta\ell_1$,
\bq\label{eq:Clestimatoraa}
\hat{C}(\ell_1) = \frac{1}{\Omega_\W} \int_{\Omega_{\ell_1}} \frac{{\rm d}^2\vell}{2\pi \ell_1\Delta_{\ell_1}}\tilde{\kappa}_\W(\vell)\tilde{\kappa}_\W(-\vell),
\eq
where the integral is taken over an annulus of width $\Delta_{\ell_1}$ centered at $\ell_1$. In this case, the Gaussian covariance can be written as
\bq\label{eq:covkappa_G_aa}
\cov_\kappa^{\rm G}(\ell_i, \ell_j) = \frac{4\pi}{\Omega_\W \ell_i \Delta_{\ell,i}} \Bigg[C(\ell_i) + \frac{\sigma_e^2}{2\bar{n}_{\rm gal}}\Bigg]^2 \delta_{ij},
\eq
where the Kronecker delta $\delta_{ij}$ ensures that the Gaussian term contributes only to the diagonal of the covariance matrix. Note, we also take into account the contribution from shot noise in shape measurements, which is given by $C^{\rm noise}= \sigma_e^2/(2\bar{n}_{\rm gal})$, where $\bar{n}_{\rm gal}= 30\ {\rm arcmin}^{-2}$ is the projected source galaxy number density and $\sigma_e=0.37$ is the RMS ellipticity of the source galaxies. We use $\ell$ bins equally spaced in log-scale with $\Delta\log\ell = 0.12$.

We evaluate the cNG contribution using Eq.~(\ref{eq:covkappa_NG}), with the response-based calculation of the angle-averaged $T_m^{\rm cNG}$ presented in Ref.~\cite{responses2}, which includes resummed tree and 1-loop level contributions. By comparing to the simulation-based estimates presented in Ref.~\cite{blot2015}, the calculation was shown to capture completely the contribution in the squeezed-regime $k_{\ell_1} \ll k_{\ell_2}$, and to account for $\approx 60\%-70\%$ for general $k_{\ell_1}, k_{\ell_2} \gtrsim 0.3 \kunit$. The difference to the treatment in Ref.~\cite{responses2} is that, in the covariance of the angle-averaged lensing power spectrum, the angle-averages are two-dimensional, and not three-dimensional. We skip writing the expressions for the projected cNG terms in this paper, as they follow straightforwardly from \refeq{limberapprox_T}, and the focus of this paper is the SSC contribution. For completeness, we note that both the monopole cNG and SSC terms do not depend on the bin width $\Delta_{\ell_1}$, contrary to the Gaussian term.

\subsection{Flat-sky vs. curved-sky SSC}\label{sec:vsfsky}

\begin{figure}
        \centering
        \includegraphics[width=\textwidth]{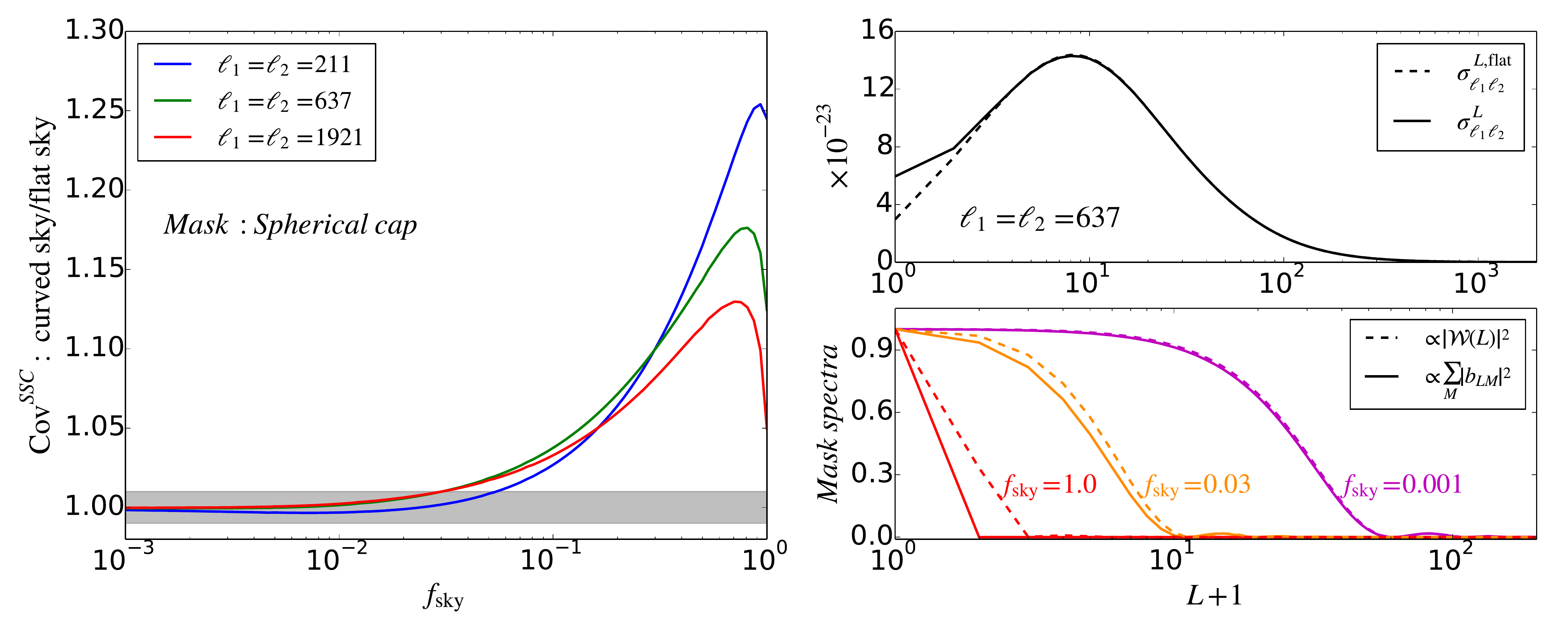}
        \caption{Comparison between the flat-sky and curved-sky SSC expressions derived in Secs.~\ref{sec:lenscovderiv_1} and \ref{sec:lenscovderiv_2}, respectively. The left panel shows the ratio of the two results as a function of the survey sky fraction $f_{\rm sky}$ covered by a spherical cap and for a few values of $\ell_1=\ell_2$, as labeled (the $f_{\rm sky}$ dependence for $\ell_1 \neq \ell_2$ is very similar to the one shown here). The gray shaded region indicates a fractional difference of 1\%. The upper right panel shows the $L$ dependence of the quantities $\sigma_{\ell_1, \ell_2}^{L, {\rm flat}}$ in Eq.~(\ref{eq:covkappa_SSC_mono_flat}) and $\sigma_{\ell_1, \ell_2}^{L}$ in Eq.~(\ref{eq:ssc_bL_1}) for one of the $\ell_1=\ell_2$ pairs shown on the left. The lower right panel shows the $L$ dependence of the flat-sky and curved-sky mask power spectra for a few $f_{\rm sky}$ values, as labeled. In the right panels, the flat-sky expressions are shown at discrete $L$ values, even though in the corresponding integral calculations $L$ varies continuously.
}
\label{fig:fsky1}
\end{figure}

To evaluate the SSC terms we consider the following two-dimensional window function 
\bq\label{eq:circlewindow}
\W(\vtheta) = 
\begin{cases}
1\ &,\  {\rm if}\ \vtheta \text{ within spherical cap with } \Omega_{\W} = f_{\rm sky} 4\pi\\
0\ &,\  {\rm otherwise}.
\end{cases}
\eq
For the flat-sky and Limber SSC result derived in Sec.~\ref{sec:lenscovderiv_1}, which we dub as {\it flat-sky} for short below, this corresponds to a disk-shape geometry with radius $\theta_\W = \sqrt{\Omega_\W/\pi}$ and for which
\bq\label{eq:circlefourier}
|\tilde{\W}(\vell)|^2 \equiv |\tilde{\W}(\ell)|^2 = \Omega_\W^2\left[\frac{2J_1(\ell\theta_\W)}{\ell\theta_\W}\right]^2,
\eq
where $J_1(x)$ is the first order Bessel function of the first kind. For the curved-sky and beyond Limber (in the super survey mode) expressions of Sec.~\ref{sec:lenscovderiv_2}, which we dub as {\it curved-sky}, we evaluate the power spectrum of the mask ($\propto \sum_M\left|b_{LM}\right|^2$) using {\sc Healpix}\footnote{\url{http://healpix.sf.net}} sphere pixelization \cite{2005ApJ...622..759G}. We choose to adopt this geometry for simplicity of calculation and because it is sufficient to illustrate the size of the anisotropic contributions in the SSC term and the difference between the flat-sky and curved-sky derivations. Naturally, this window function shape may not be representative of the general anisotropic and irregular masks that characterize real surveys. We leave the exploration of more general window functions for future work.

The left panel of Fig.~\ref{fig:fsky1} compares the curved-sky and flat-sky SSC results. As expected, the two results approach one another with decreasing $f_{\rm sky}$ values. However, while for surveys covering less than $\approx 1\%-5\%$ of the sky the two results agree to better than $1\%$, indicated by the gray shaded area, for surveys such as Euclid \cite{2011arXiv1110.3193L} or LSST \cite{2012arXiv1211.0310L} with $f_{\rm sky} \approx 0.3-0.4$ the use of the flat-sky expressions results in an underestimation of the SSC contribution of about $10\%$. For accurate error estimates, one should thus use the curved-sky result of Eq.~(\ref{eq:ssc_bL_1}).

The right panels in Fig.~\ref{fig:fsky1} help to understand the result shown on the left. First, the upper panel shows that $\sigma_{\ell_1, \ell_2}^{L}$ and $\sigma_{\ell_1, \ell_2}^{L, {\rm flat}}$ differ only at low $L$. Second, the lower panel shows that larger $f_{\rm sky}$ values up-weight the contribution from low $L$ values in the integral of Eq.(\ref{eq:covkappa_SSC_mono_flat}) and sum of Eq.~(\ref{eq:ssc_bL_1}). Hence, for large enough $f_{\rm sky}$, most of the contribution comes from low $L$, which is where $\sigma_{\ell_1, \ell_2}^{L} \neq \sigma_{\ell_1, \ell_2}^{L, {\rm flat}}$ and $\left|\W(L)\right|^2 \neq \sum_M\left|b_{LM}\right|^2$, thus explaining why the resulting SSC values differ as well. On the other hand, lower values of $f_{\rm sky}$ up-weight the contribution from larger $L$ values where $\sigma_{\ell_1, \ell_2}^{L} \approx \sigma_{\ell_1, \ell_2}^{L, {\rm flat}}$ and $\left|\W(L)\right|^2 \approx \sum_M\left|b_{LM}\right|^2$, effectively bringing the two results together. The turnover at $f_{\rm sky} \approx 0.8$ reflects the $f_{\rm sky}$ dependence of the competition between the beyond Limber corrections that work to enhance the covariance ($\sigma_{\ell_1, \ell_2}^L > \sigma_{\ell_1, \ell_2}^{L, {\rm flat}}$) and the beyond flat-sky corrections in the mask power spectrum that work to suppress it ($\sum_M\left|b_{LM}\right|^2 < \left|\W(L)\right|^2$).

In Fig.~\ref{fig:fsky1} we only show results for $\ell_1 = \ell_2$ for brevity, but we have explicitly checked that the same conclusions hold for the off-diagonal ($\ell_1 \neq \ell_2$) cases as well.

\subsection{SSC vs. cNG vs. G}\label{sec:lenscovresults}

We now turn our attention to a comparison between the size of the Gaussian (G), connected non-Gaussian (cNG) and super-sample (SSC) contributions to the covariance of the monopole lensing convergence power spectrum. In this subsection, we consider only a single survey sky fraction, $\Omega_{\W} = f_{\rm sky}4\pi \approx 15000\ {\rm deg}^2$ ($f_{\rm sky} \approx 0.36$), and the SSC results correspond to the curved-sky derivation of Sec.~\ref{sec:lenscovderiv_2}.

\begin{figure}
        \centering
        \includegraphics[width=\textwidth]{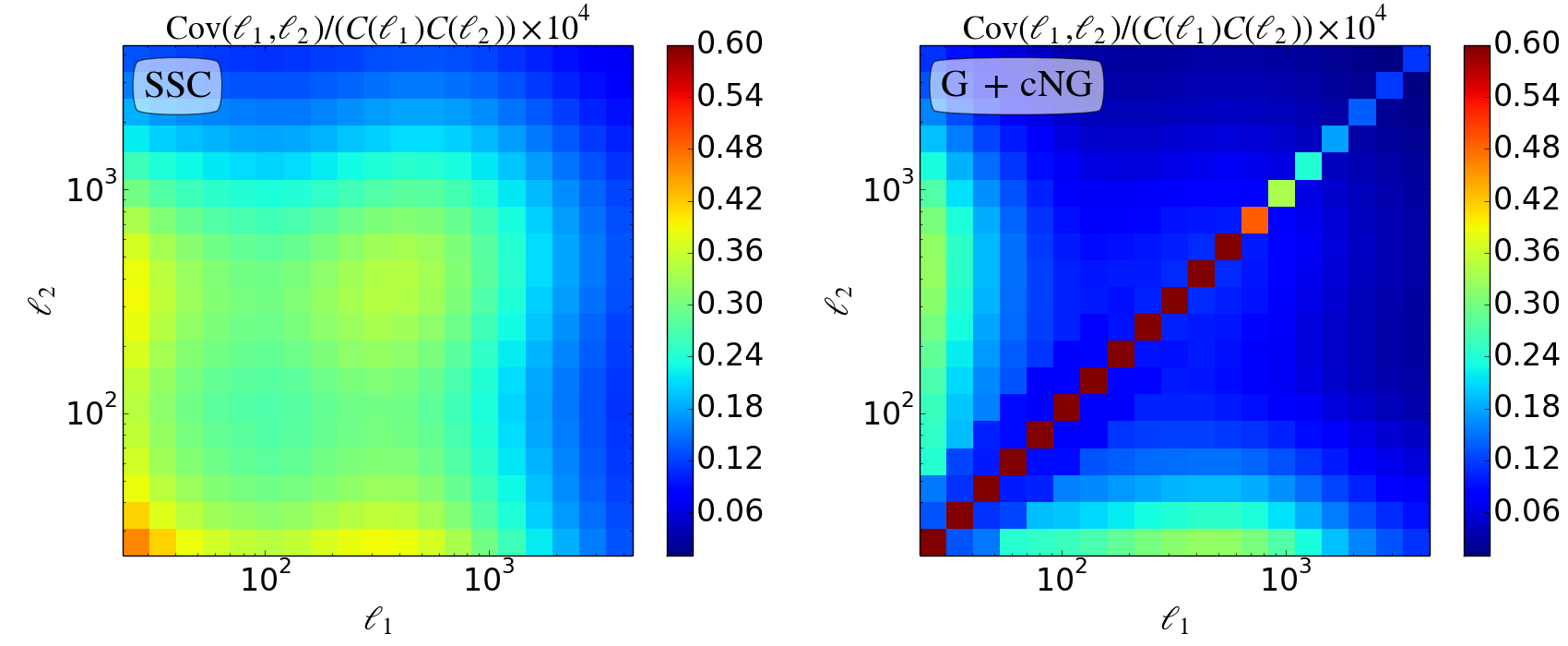}
        \caption{Contributions to the angle-averaged (monopole) lensing convergence covariance matrix. The left color map shows the curved-sky SSC covariance contribution computed with Eq.~(\ref{eq:ssc_bL_1}) for a spherical cap with $f_{\rm sky} = 0.36$ as mask. The right color map shows the same, but for the sum of the Gaussian (cf.~Eq.~(\ref{eq:covkappa_G_aa})) and connected non-Gaussian contributions (cf.~Eq.~(\ref{eq:covkappa_NG}) with the response-based calculation of Ref.~\cite{responses2}). In the right panel, the diagonal terms up to $\ell_1, \ell_2 \sim 500$ are higher than the maximum of the color scale (i.e., the color scale saturates) to facilitate visualization of the off-diagonal terms.}
\label{fig:maps_isowindow}
\end{figure}

\begin{figure}
        \centering
        \includegraphics[width=\textwidth]{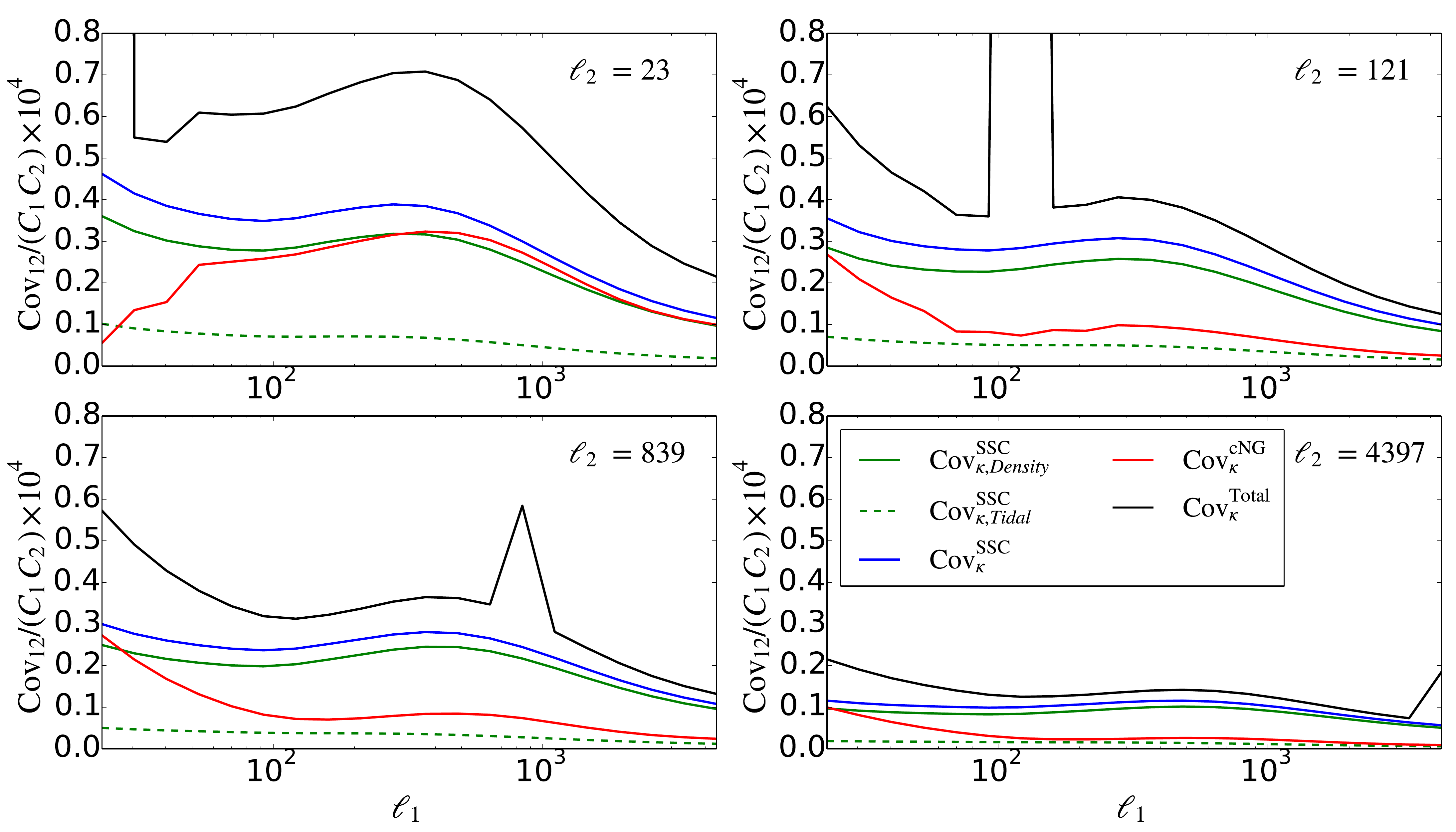}
        \caption{Angle-averaged (monopole) lensing convergence covariance matrix as a function of $\ell_1$, for fixed values of $\ell_2$, as labeled. The blue line shows the SSC term, the solid green line shows the same, but assuming $R_K = 0$ (i.e., considers only super-survey isotropic density fluctuations); the dashed green line shows the difference between the two and thus measures the contribution from super-survey tidal fields. The red line shows the connected non-Gaussian term computed using Eq.~(\ref{eq:covkappa_NG}) with the response calculation of Ref.~\cite{responses2}. The black solid line shows the total lensing covariance, including the Gaussian contribution. In the labels, $\cov_{12} = \cov(\ell_1, \ell_2)$ and $C_i = C(\ell_i)$.}
\label{fig:slices_isowindow}
\end{figure}

\begin{figure}
        \centering
        \includegraphics[width=\textwidth]{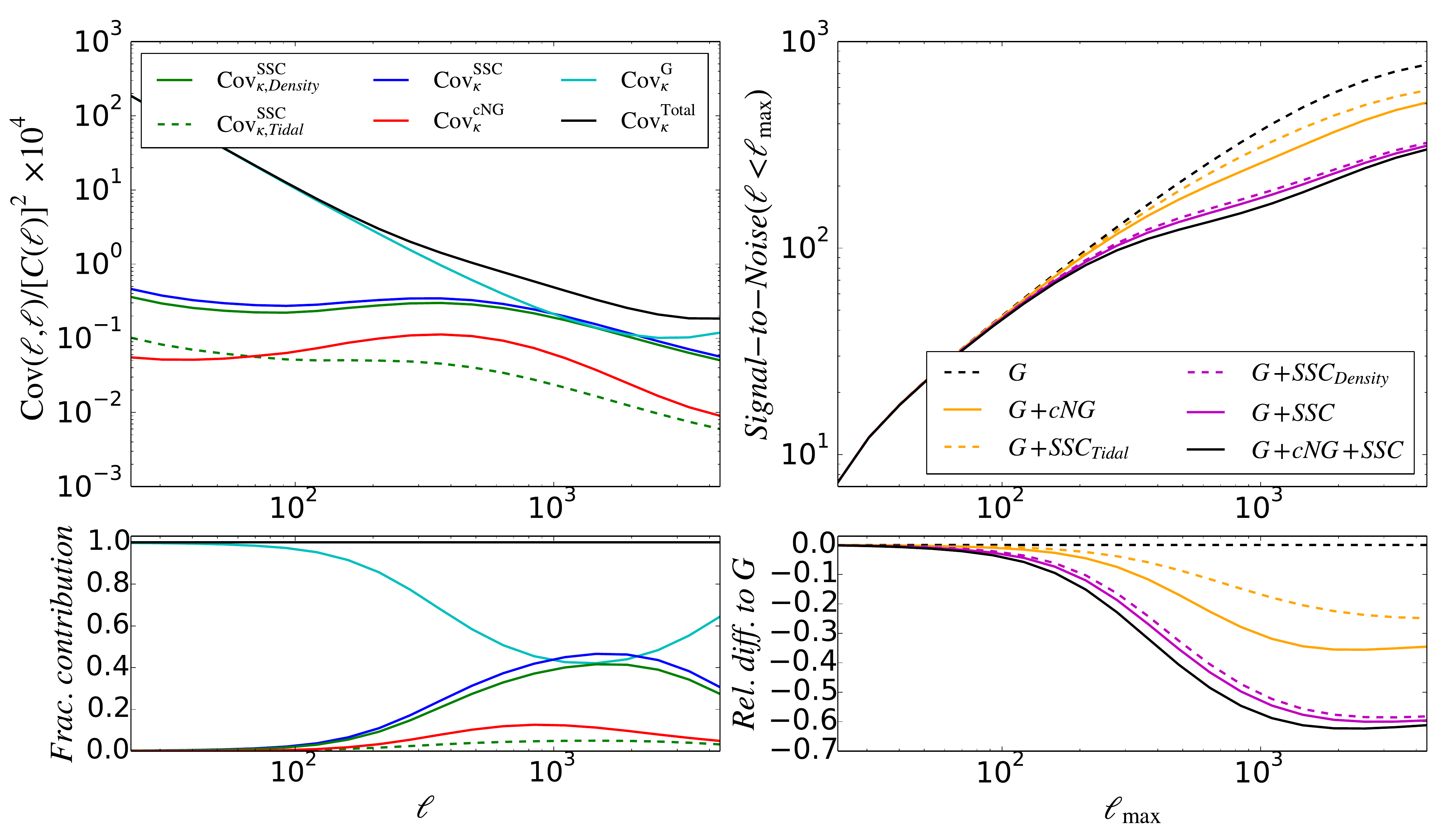}
        \caption{The upper left panel shows the same as Fig.~\ref{fig:slices_isowindow}, but along the diagonal of the covariance matrix, $\ell_1=\ell_2$; the lower left panel shows the fractional size of each contribution to the total covariance. The upper right panel shows the signal-to-noise ratio as a function of $\ell_{\rm max}$ (cf.~Eq.~(\ref{eq:sn})); the lower right panel shows the relative difference to the Gaussian-only case.
       }
\label{fig:diagonal_isowindow}
\end{figure}

The color maps in Fig.~\ref{fig:maps_isowindow} show the SSC (left), and the sum of the G and cNG (right) contributions to the total covariance. Figure \ref{fig:slices_isowindow} shows a few {\it slices} of the various contributions at constant $\ell_2$, as labeled. The two left panels of Fig.~\ref{fig:diagonal_isowindow} show the same, but along the diagonal $\ell_1=\ell_2$. Finally, the two right panels of Fig.~\ref{fig:diagonal_isowindow} show the signal-to-noise ratio defined as
\bq\label{eq:sn}
\left(\frac{S}{N}\right)^2_{< l_{\rm max}} = \sum_{\ell_1, \ell_2 < \ell_{\rm max}} C(\ell_1)\cov^{-1}(\ell_1, \ell_2)C(\ell_2),
\eq
as a function of the maximum wavenumber $\ell_{\rm max}$. This helps to quantify the loss of information caused by the various contributions to the covariance matrix.

Figures \ref{fig:maps_isowindow}, \ref{fig:slices_isowindow} and \ref{fig:diagonal_isowindow} illustrate the following main points:

\begin{enumerate}
\item The SSC is the dominant off-diagonal contribution. For Euclid- and LSST-like surveys, which will observe about $30-40\%$ of the sky, these results suggest that the SSC contribution constitutes a significant portion ($\approx 40-50\%$ for $\ell \approx 10^3$ along the diagonal) of the statistical error that should not be ignored.

\item The summed contribution from the tidal SSC terms derived in this paper (those proportional to $R_K$; dashed green lines labeled as $SSC_{\rm tidal}$), although non-negligible, is appreciably smaller than the contribution from the purely isotropic density SSC term (labeled as $SSC_{\rm density}$). For instance, on the diagonal (cf.~left panels of Fig.~\ref{fig:diagonal_isowindow}), these terms constitute only $\approx 5\%$ of the total covariance matrix at $\ell \gtrsim 300$.

\item In terms of signal-to-noise, the degradation relative to the purely Gaussian case is strongly dominated by the SSC contribution, as shown by the similarity between the $G+SSC$ (solid magenta) and $G + cNG + SSC$ (solid black) curves in the right panel of \reffig{diagonal_isowindow}. The SSC contribution is in turn dominated by the $SSC_{\rm density}$ term, as shown by the similarity between the $S+SSC_{\rm density}$ (dashed magenta) and $G+SSC$ (solid magenta) curves.

\end{enumerate}

The conclusions drawn above should be interpreted in light of the rather simplistic survey setup adopted. The exact relative contribution of the various terms  can depend on the shape of the source galaxy redshift distribution and number of tomographic bins used, as well as on the shape of the window function for the case of the SSC terms \cite{2014MNRAS.444.3473T, 2016arXiv161205958L}. We leave such developments for a future dedicated forecast study where we will aim to quantify the importance of the various covariance contributions for future surveys (e.g.~Euclid \cite{2011arXiv1110.3193L} and LSST \cite{2012arXiv1211.0310L}).

The impact of the window function on the cNG term is also ignored in our results \cite{2004A&A...413..465K,2004MNRAS.349..603E}, but we note that this is not expected to have any significant impact. For instance, in Ref.~\cite{2014MNRAS.444.3473T}, the authors show that the covariance estimated from masked log-normal lensing maps with $100\ {\rm deg}^2$ agrees very well with an analytical derivation that only considers window function effects in the G and SSC term; this indicates that window function effects in the cNG term (not counting the scaling with the survey area) can be safely ignored\footnote{The impact of the window function in the G term can be included by considering the windowed power spectra in Eq.~(\ref{eq:covkappa_G_aa}), instead of the expectation value. However, in the $\ell \gg 1/\theta_0$ limit we considered here, these two spectra are the same (cf.~comment after Eq.~(\ref{eq:Clestimator})) and hence we can skip using the windowed power spectrum explicitly.}.  A drawback that is normally associated with analytical covariance estimates concerns precisely the difficulties to incorporate window function effects, which is arguably true for the case of the cNG term. However, if these window function effects are negligible compared to the most dominant effect in the SSC term (which is straightforwardly calculable with the knowledge of $\W(\vtheta)$), then this drawback of analytical approaches becomes irrelevant in practice.

It is also worth recalling that all our results assume the Limber approximation for the sub-survey modes. For the G term, the validity of Limber's approximation is the same as that for the power spectrum, and hence, one is justified in adopting it for $\ell \gtrsim 20$. In our curved-sky derivation of the SSC term, the starting point was to replace the power spectrum $P_m(k,z)$ by the power spectrum modulated by long-wavelength perturbations $P_m(k, z | \vx) - P_m(k,z) = \left[R_1(k,z)\delta(\vx) + R_K(k,z)\hat{\vk}^i\hat{\vk}^jK_{ij}(\vx)\right]$ in Eq.~(\ref{eq:Cl_pi_1}). In this respect, the angular modes $\ell$ continue to get contributions from essentially the same range of $k$ values (which now appear also in the arguments of the response functions), and hence, the adoption of Limber's approximation should remain as valid as for the power spectrum. A robust assessment of the size of beyond flat-sky and Limber's approximation terms in the cNG term is appreciably harder to carry out as it (to the best of our knowledge) involves dealing with high-dimensional integrals of the fully general unequal-time matter trispectrum (cf. Ref.~\cite{2017arXiv171107372L} and Appendix \ref{app:general}); it could presumably be easier to make comparisons against purely numerical estimates of the lensing non-Gaussian covariance obtained with full-sky lensing simulations (e.g.~Ref.~\cite{2018arXiv180105745S})\footnote{A potential problem with measuring the cNG term using such simulations is that there will be modes inside the box that are outside the simulated lightcone, and hence, induce some SSC contributions. The severity of this problem can nonetheless be reduced by using smaller size cubic boxes to simulate regions of the lightcone closer to the observer.}. Given however that no survey-scale modes are involved in the cNG contribution, we expect the difference between the curved- and flat-sky results to be significantly smaller for the cNG case, compared to SSC. Hence, althought we do not present a rigorous proof (and we are not aware of any in the literature), we roughly expect the flat-sky cNG formulae used here to represent a good approximation for use in real survey analyses on scales $\ell \gtrsim 20$, i.e., scales for which the Limber approximation is also valid for the power spectrum. One can further argue that the exact validity of flat-sky and Limber approximations in the cNG term is not as worrying as it could be for the SSC case because of its small contribution to the total covariance; for instance, a change of $\approx 10\%$ in the amplitude of the cNG term (which is approximately the size of the differences between the flat- and curved-sky SSC results for $f_{\rm sky} \approx 0.3$, and thus a conservative upper limit on the corresponding effect for cNG) has barely any impact in the signal-to-noise curves depicted in Fig.~\ref{fig:diagonal_isowindow}.

\section{Summary and conclusions}\label{sec:conc}

We have used the response-based extension of perturbation theory to derive the full super-sample contribution (SSC) to the covariance of the matter, as well as lensing convergence power spectrum. The original derivation of Ref.~\cite{takada/hu:2013} described the contribution given by the response of the power spectrum to purely isotropic super-survey modes. Here, we have derived the remainder of the lensing SSC, which corresponds to terms that involve the response of the power spectrum to super-survey tidal fields. The latter contribute to the monopole lensing SSC via their projection along the line-of-sight and the trace of their projection onto the sky (the traceless part averages out); this is different than in the three-dimensional case, in which tidal fields do not contribute after angle-averaging regardless of the shape of the window function (cf.~comment after Eq.~(\ref{eq:eqmonoPw})). Further, we have presented a rigorous generalization of the SSC contribution beyond the flat-sky and Limber approximations, which is valid as long as the lensing angular power spectra are evaluated for $\ell_1,\ell_2 \gtrsim 20$ (cf.~Sec.~\ref{sec:lenscovderiv_2}). Overall, our response-based derivation proved particularly useful in clarifying the origin of all the terms that make up the SSC contribution (Sec.~\ref{sec:lenscovderiv_1}).

We have compared the flat- and curved-sky SSC results numerically (cf.~Fig.~\ref{fig:fsky1}) and found that they agree to better than $1\%$ for survey sky fractions of $f_{\rm sky} \lesssim 0.01 - 0.05$, but the differences can become of order $10\%$ for $f_{\rm sky} \approx 0.35$. It is thus recommended that the curved-sky SSC expressions (see also Ref.~\cite{2016arXiv161205958L}) are used in parameter inference analyses or forecast studies using weak lensing data from surveys with significant sky coverage, specially given that their implementation in existing inference codes is straightforward.

For an idealized lensing survey setup with $f_{\rm sky} \approx 0.36$, single source redshift at $z_S=1$ and isotropic window function, we have shown that the SSC term is the dominant off-diagonal contribution to the total lensing covariance matrix; it also becomes comparable to the diagonal Gaussian contribution for $\ell_1= \ell_2 \gtrsim 600$. Additionally, we have demonstrated that the contribution from the tidal SSC terms derived here is at the level of $\approx 5\%$ of the total lensing covariance for $\ell_1, \ell_2 \gtrsim 10^3$. The extent to which these tidal terms can actually impact parameter constraints will be studied in a follow-up forecast study with more realistic window function geometries and source redshift distributions. The results shown here provide a first indication that their contribution is expected to be smaller than that of the cNG term (cf.~right panels of Fig.~\ref{fig:diagonal_isowindow}).

The cosmological analysis of large-scale structure surveys is naturally not limited to lensing observations. In recent works within the KiDS \cite{2017arXiv170706627J, 2017arXiv170605004V} and DES \cite{2017arXiv170801530D} collaborations, the resulting parameter constraints were obtained from a joint analysis of lensing and projected galaxy two-point statistics, including their cross-correlation. Future observational efforts are likely to involve similar such combined analyses. The response formalism can also in principle be extended to galaxies as density tracers and be used to calculate the corresponding cNG and SSC terms of their auto power spectrum, as well as of the cross-spectrum with lensing. This will require separate universe measurements of the responses of the galaxy power spectrum (see e.g.~Ref.~\cite{2017arXiv171100018L} for a measurement of such response coefficients using power spectra measured in sub-volumes of a periodic $N$-body simulation box).

Finally, we note that the steps that we have taken here can be applied also to the calculation of the super-sample bispectrum covariance. The inclusion of the bispectrum in cosmological analyses can improve parameter constraints \cite{2013MNRAS.429..344K, 2006PhRvD..74b3522S, 2017PhRvD..96b3528C, 2018arXiv180206762D}, and as a result, it is important that its covariance matrix (and cross-covariance with the power spectrum) is equally well understood. In Ref.~\cite{2017arXiv170902473C}, the authors have taken the first steps in this direction, but similarly to Ref.~\cite{takada/hu:2013}, their results take into account only responses to isotropic super-survey modes computed using the halo model. Within the response formalism, the generalization to the case of the bispectrum is straightforward, but will require simulation measurements of the first order bispectrum response coefficients, which can be done with the separate universe technique.

\begin{acknowledgments}

We thank Kazuyuki Akitsu, Fabien Lacasa, Yin Li, Takahiro Nishimichi, and Masahiro Takada for useful comments and discussions. FS acknowledges support from the Starting Grant (ERC-2015-STG 678652) ``GrInflaGal'' from the European Research Council. EK acknowledges support from NASA grant 15-WFIRST15-0008 Cosmology with the High Latitude Survey WFIRST Science Investigation Team (SIT). Some of the results in this paper have been derived using the HEALPix \cite{2005ApJ...622..759G} package.

\end{acknowledgments}

\appendix

\section{Derivation of the three-dimensional matter covariance with a finite window}\label{app:covderiv}

In this appendix, we show a few of the intermediate steps taken when deriving Eq.~(\ref{eq:covres}) from Eq.~(\ref{eq:covdef}). The following is an identity that will be useful below:
\bq
\label{eq:useful1} \tilde{W}(\vk) = \int_{\vp}\tilde{W}(\vp)\tilde{W}(\vk-\vp) = \left[\prod_{i=1}^n \int_{\vp_i}\tilde{W}(\vp_i)\right](2\pi)^3\delta_D(\vk - \vp_{12..n}),
\eq
which follows from the fact that the window function in real space satisfies $W^n(\vx)=W(\vx)$.

Rewriting Eq.~(\ref{eq:covdef}) here, we have
\bq\label{eq:covdefapp}
\cov(\vk_1, \vk_2)&=& \langle\hat{P}_W(\vk_1)\hat{P}_W(\vk_2)\rangle - \langle\hat{P}_W(\vk_1)\rangle \langle\hat{P}_W(\vk_2)\rangle \nonumber \\
&=& \frac{1}{V_W^2}\Big[\langle\tilde{\delta}_W(\vk_1)\tilde{\delta}_W(\vk_2)\rangle\langle\tilde{\delta}_W(-\vk_1)\tilde{\delta}_W(-\vk_2)\rangle \ + \  \left(\vk_2 \leftrightarrow -\vk_2 \right)\Big] \nonumber \\
&+& \frac{1}{V_W^2}\langle\tilde{\delta}_W(\vk_1)\tilde{\delta}_W(-\vk_1)\tilde{\delta}_W(\vk_2)\tilde{\delta}_W(-\vk_2)\rangle_c.
\eq
The terms inside square brackets in Eq.~(\ref{eq:covdefapp}) can be worked out as
\bq\label{eq:gaussderiv}
&&\langle\tilde{\delta}_W(\vk_1)\tilde{\delta}_W(\vk_2)\rangle\langle\tilde{\delta}_W(-\vk_1)\tilde{\delta}_W(-\vk_2)\rangle = \nonumber \\
&=& \left[\prod_{i=1}^4 \int_{\vp_i} \tilde{W}(\vp_i)\right] (2\pi)^3 \delta_D(\vk_{12} - \vp_{12}) (2\pi)^3 \delta_D(-\vk_{12} - \vp_{34}) \Pnl(\vk_1-\vp_1)\Pnl(\vk_1+\vp_3) \nonumber \\
&=& [\Pnl(\vk_1)]^2  \int_{\vp_1}\int_{\vp_2}\int_{\vp_3} \tilde{W}(\vp_1)\tilde{W}(\vp_2)\tilde{W}(\vp_3)\tilde{W}(-\vk_{12}-\vp_3) (2\pi)^3 \delta_D(\vk_{12} - \vp_{12}) \nonumber \\
&=& [\Pnl(\vk_1)]^2  \int_{\vp_3} \tilde{W}(\vk_{12})\tilde{W}(\vp_3)\tilde{W}(-\vk_{12}-\vp_3) \nonumber \\
&=& [\Pnl(\vk_1)]^2  |\tilde{W}(\vk_{12})|^2.
\eq
In the first equality above, we used Eq.~(\ref{eq:dfourier}) and the definition of the power spectrum; in the second equality, the integral over ${\vp_4}$ eliminates one Dirac-delta and the approximation $\Pnl(\vk_1+\vp_i) \approx \Pnl(\vk_1)$ allows us to move the power spectra terms out of the integrals; in the third equality we used Eq.~(\ref{eq:useful1}) for $n=2$, and then again in the fourth equality. This result is that of the first term on the right-hand side of Eq.~(\ref{eq:covres}).

The derivation of the connected four-point function term in Eq.~(\ref{eq:covdefapp}) can be done as follows
\bq\label{eq:ngaussderiv}
&&\langle\tilde{\delta}_W(\vk_1)\tilde{\delta}_W(-\vk_1)\tilde{\delta}_W(\vk_2)\tilde{\delta}_W(-\vk_2)\rangle_c = \nonumber\\
&=& \left[\prod_{i=1}^4 \int_{\vp_i} \tilde{W}(\vp_i)\right] (2\pi)^3 \delta_D(\vp_{1234}) T_m(\vk_1-\vp_1, -\vk_1-\vp_2, \vk_2-\vp_3, -\vk_2-\vp_4) \nonumber \\
&=& \left[\prod_{i=1}^4 \int_{\vp_i} \tilde{W}(\vp_i)\right] (2\pi)^3 \delta_D(\vp_{1234}) T_m(\vk_1, -\vk_1-\vp_{12}, \vk_2, -\vk_2-\vp_{34}) \nonumber \\
&=&  \int_{\vp_1} \int_{\vp_2} \int_{\vp_3} \tilde{W}(\vp_1)\tilde{W}(\vp_2)\tilde{W}(\vp_3)\tilde{W}(-\vp_{123}) T_m(\vk_1, -\vk_1-\vp_{12}, \vk_2, -\vk_2+\vp_{12}).
\eq
The first equality again simply uses the definition of the matter trispectrum and of the observed density contrast Eq.~(\ref{eq:dfourier}). In the second equality, we have performed the change of variables $\vk_1 - \vp_1 \to \vk_1$, $\vk_2 - \vp_3 \to \vk_2$, which is allowed under the assumed limit that $k_i \gg p_i$; in the third equality, we have integrated over $\vp_4$. From hereon, the steps involve manipulations of the integrals over the window functions. Defining a new integration variable $\vp = \vp_{12}$, Eq.~(\ref{eq:ngaussderiv}) continues as
\bq\label{eq:ngaussderiv2}
&&\int_{\vp_1} \int_{\vp} \int_{\vp_3} \tilde{W}(\vp_1)\tilde{W}(\vp-\vp_1)\tilde{W}(\vp_3)\tilde{W}(-\vp - \vp_3) T_m(\vk_1, -\vk_1-\vp, \vk_2, -\vk_2+\vp) \nonumber \\
&=&\int_{\vp}|W(\vp)|^2 T_m(\vk_1, -\vk_1+\vp, \vk_2, -\vk_2-\vp),
\eq
where we have used Eq.~(\ref{eq:useful1}) for $n=1$ twice in the integrals over ${\vp_1}$ and ${\vp_3}$. This matches the result of Eq.~(\ref{eq:covres}).

\section{The super sample term at tree level}\label{app:ssctree}

To build intuition about the origin of the super-sample term derived in Sec.~\ref{sec:ssc}, it is instructive to understand how it arises at tree level in perturbation theory (see also Appendix A of Ref.~\cite{2009MNRAS.395.2065T}). The tree-level matter trispectrum is given by
\bq\label{eq:treetrispectrum}
T_m(\vk_a, \vk_b, \vk_c, \vk_d) &=& \Big[6\fiii(\vk_a, \vk_b, \vk_c)\Plin(k_a)\Plin(k_b)\Plin(k_c) + 3\ {\rm perm.}\Big] \nonumber \\
&+& \Big[ 4\fii(-\vk_{ab}, \vk_b)\fii(\vk_{ab}, \vk_c) \Plin(|\vk_{ab}|)\Plin(k_b) \Plin(k_c) + 11 \ {\rm perm.} \Big],
\eq
where $\fii$ and $\fiii$ are the second- and third-order perturbation theory kernels. By specializing to the relevant configuration, it is straightforward to show that (always implicitly assuming $p \ll k_1, k_2$)
\bq\label{eq:treessc_dec}
T_m(\vk_1, -\vk_1 + \vp, \vk_2, -\vk_2 - \vp) \approx \underbrace{T_m^{\rm cNG,tree}(\vk_1, -\vk_1, \vk_2, -\vk_2)}_{4 \fiii\ {\rm terms}\ +\ 8\fii^2\ {\rm terms}} + \underbrace{T_m^{\rm SSC, tree}(\vk_1, -\vk_1, \vk_2, -\vk_2; \vp)}_{4 \fii^2\  {\rm terms}}. \nonumber \\
\eq
That is, out of the 16 permutations in Eq.~(\ref{eq:treetrispectrum}), there are 12 that asymptote to a finite value as $p \to 0$. These correspond to the standard connected non-Gaussian tree-level covariance (see e.g. Eq.~(3.1) of Ref.~\cite{responses2}). The remaining four permutations of Eq.~(\ref{eq:treetrispectrum}) are $\propto \Plin(p)$, and are thus precisely the terms isolated by the limit in \refeq{Tssc}. These therefore represent the SSC term. Diagrammatically, at tree level we have
\bq\label{eq:sscdiagrams_tree}
T^{\rm SSC, tree}(\vk_1, -\vk_1, \vk_2, -\vk_2; \vp) &=& \left[\raisebox{-0.0cm}{\includegraphicsbox[scale=0.8]{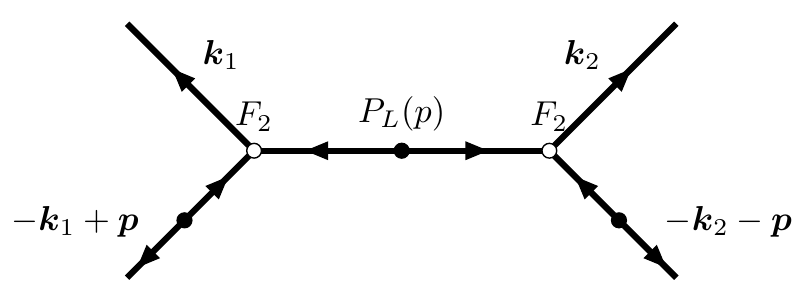}} + (\vk_1 \leftrightarrow -\vk_1+\vp)\right] \nonumber \\ 
&&+ (\vk_2 \leftrightarrow -\vk_2-\vp) \nonumber \\ 
&=& \Big[2\fii(\vk_1-\vp, \vp)\Plin(|\vk_1-\vp|) + 2\fii(\vk_1, -\vp)\Plin(k_1)\Big]   \nonumber \\
&\times& \Big[2\fii(\vk_2+\vp, -\vp)\Plin(|\vk_2+\vp|) + 2\fii(\vk_2, \vp)\Plin(k_2)\Big] \Plin(p) \nonumber \\
&=& \R^{\rm tree}_1(k_1, - \mu_{\vp,\vk_1}) \R^{\rm tree}_1(k_2, \mu_{\vp,\vk_2})\Plin(k_1)\Plin(k_2)\Plin(p),
\eq
where in the last equality we have used the tree level limit of the first order response $\R_1$ derived in Ref.~\cite{responses1}. The expression above is only valid if all modes are in the linear regime $k_1, k_2, p \ll \knl$. However, the resummed response vertices, obtained using simulation measurements \cite{li/hu/takada, response, andreas} of the first order response $R_1,\,R_K$ (cf.~Fig.~\ref{fig:resp_zevo}), allow us to be predictive for $p \ll \knl$, but any nonlinear $k_1$, $k_2$.

\section{The Limber approximation in the flat-sky lensing covariance}\label{app:limber}

In this appendix, we derive the Limber-approximated lensing convergence power spectrum and trispectrum expressions that we used in Sec.~\ref{sec:lenscovderiv_1} (cf.~Eqs.~(\ref{eq:limberapprox_C}) and (\ref{eq:limberapprox_T})).

The lensing convergence power spectrum is defined in the flat-sky limit as $(2\pi)^2\delta_D(\vell + \vell')C(\vell) = \langle\tilde{\kappa}(\vell)\tilde{\kappa}(\vell')\rangle$. Integrating over ${\rm d}^2\vell'$, the power spectrum can be worked out as follows:
\bq\label{eq:Cllimber}
C(\vell) &=& \int \frac{{\rm d^2\vell'}}{(2\pi)^2} \langle\tilde{\kappa}(\vell)\tilde{\kappa}(\vell')\rangle \nonumber \\
&=& \int \frac{{\rm d^2\vell'}}{(2\pi)^2} \int {\rm d}^2\vtheta_1{\rm d}^2\vtheta_2 \langle\kappa(\theta_1)\kappa(\theta_2)\rangle e^{i\vell \vtheta_1}e^{i\vell' \vtheta_2}  \nonumber \\
&=& \int \frac{{\rm d^2\vell'}}{(2\pi)^2} \int {\rm d}^2\vtheta_1{\rm d}^2\vtheta_2 \int {\rm d}\chi_1{\rm d}\chi_2\ g(\chi_1) g(\chi_2) \langle\delta(\chi_1\vtheta_1, \chi_1)\delta(\chi_2\vtheta_2, \chi_2)\rangle e^{i\vell \vtheta_1}e^{i\vell' \vtheta_2} \nonumber\\
&=& \int \frac{{\rm d^2\vell'}}{(2\pi)^2} \int {\rm d}^2\vtheta_1{\rm d}^2\vtheta_2\int  {\rm d}\chi_1{\rm d}\chi_2 g(\chi_1) g(\chi_2) \int \frac{{\rm d}^3\vk}{(2\pi)^3} \Pnl(\vk, \chi_1, \chi_2) \nonumber \\ 
&& \ \ \ \ \ \ \ \ \ \ \ \ \ \ \ \ \ \ \ \ \ \ \ \ \ \ \ \ \ \ \ \ \ \ \ \ \ \ \times e^{-i\vk_{\perp}(\chi_1\vtheta_1 - \chi_2\vtheta_2)} e^{-ik_\parallel(\chi_1 - \chi_2)} e^{i\vell \vtheta_1}e^{i\vell' \vtheta_2} \nonumber\\
&=& \int {\rm d}^2\vtheta_1 \int {\rm d}\chi_1{\rm d}\chi_2\ g(\chi_1) g(\chi_2) \int \frac{{\rm d}^3\vk}{(2\pi)^3} \Pnl(\vk, \chi_1, \chi_2) e^{-ik_\parallel(\chi_1 - \chi_2)} e^{i\vtheta_1(\vell-\vk_{\perp}\chi_1)}\nonumber \\ 
&=& \int  {\rm d}\chi_1{\rm d}\chi_2 \frac{g(\chi_1)}{\chi_1^2} g(\chi_2) \int \frac{{\rm d}k_\parallel}{2\pi} \Pnl\left(\vk=\left(\frac{\vell}{\chi_1}, k_\parallel\right), \chi_1, \chi_2\right) e^{-ik_\parallel(\chi_1 - \chi_2)}\nonumber \\ 
\eq
In the second equality we have written down the inverse Fourier transformed $\tilde{\kappa}$; in the third equality we have used the definition of the convergence as a weighted density projection, Eq.~(\ref{eq:kappadef}); in the fourth equality we have expanded the density in real space into Fourier modes, used the definition of the unequal-time matter power spectrum $(2\pi)^3\delta(\vk + \vk')\Pnl(\vk, \chi_1, \chi_2)$ $=\langle\delta(\vk, \chi_1)\delta(\vk', \chi_2)\rangle$, and integrated over ${\rm d}^3\vk'$ using the Dirac-delta function; in the fifth equality, we used that the integration over ${\rm d}^2\vell'$ yields $\delta_D(\vtheta_2)$ to integrate over ${\rm d}^2\vtheta_2$; finally, in the last equality, we used that the integration over ${\rm d}^2\vtheta_1$ yields a Dirac-delta $\delta_D(\vell-\vk_{\perp}\chi_1)$ to integrate over ${\rm d}^2\vk_\perp$, which sets $\vk_\perp = \vell/\chi_1$ with $\vk = \left( \vk_\perp, k_\parallel \right)$.

The argumentation behind Limber's approximation goes as follows. If one is interested in sufficiently small angular scales (large $\ell$), then the dependence of the power spectrum on $k_\parallel$ can be neglected. This is because the contribution from modes $k_\parallel \gtrsim 1/\chi$ (where $\chi$ here denotes the typical line-of-sight distances involved) is suppressed by the oscillations of the integrand $\propto e^{-ik_\parallel(\chi_1 - \chi_2)}$. If we are then interested in cases of large $\ell$, then $\ell/\chi \gg k_\parallel$ and we can set $\vk \approx (\vell/\chi, 0)$ in the argument of the power spectrum in Eq.~(\ref{eq:Cllimber}). Doing so, the integral over ${\rm d}k_\parallel$ yields a Dirac-delta $\delta_D(\chi_1-\chi_2)$, which allows to arrive at Eq.~(\ref{eq:limberapprox_C}):
\bq
C(\vell) &=& \int{\rm d}\chi [g(\chi)]^2 \chi^{-2} \Pnl\left(\vk = \frac{\vell}{\chi}; z(\chi)\right).
\eq
As we will see in the next appendix (and as we have used in the main body of the paper), the result can be made slightly more precise if $\vk \to \vk_{\ell} = \hat{\vell} (\ell + 1/2)/\chi$; this correction is nonetheless unimportant for $\ell \gg 1$, as it was assumed.

For the case of the trispectrum, the derivation is analogous to that of the power spectrum, just with a larger number of integrals. In the flat-sky limit, the convergence trispectrum can be written as 
\bq
(2\pi)^2\delta_D(\vell_{abcd})\T_\kappa(\vell_a,\vell_b,\vell_c,\vell_d) = \langle\tilde{\kappa}(\vell_a)\tilde{\kappa}(\vell_b)\tilde{\kappa}(\vell_c)\tilde{\kappa}(\vell_d)\rangle,
\eq
from which it follows that
\bq
\T(\vell_a,\vell_b,\vell_c,-\vell_{abc}) &=& \int \frac{{\rm d}^2\vell_d}{(2\pi)^2}\Bigg[\prod_{m=a,b,c,d}\int{\rm d}^2\vtheta_m\int{\rm d}\chi_m g(\chi_m)e^{i\vell_m\vtheta_m}\int\frac{{\rm d}^3\vk_m}{(2\pi)^{12}} \nonumber \\
&& \ \ \ \ \ \ \ \ \ \ \ \ \ \ \ \ \ \ \ \ \ \ \times e^{-i\vk_{m,\perp}\vtheta_m\chi_m}e^{-ik_{m,\parallel}\chi_m}\Bigg] \nonumber \\
&&\ \ \ \ \ \ \ \ \ \ \  \times \langle\tilde{\delta}(\vk_a, \chi_a)\tilde{\delta}(\vk_b, \chi_b)\tilde{\delta}(\vk_c, \chi_c)\tilde{\delta}(\vk_d, \chi_d)\rangle.
\eq
Using the definition of the three-dimensional matter trispectrum, integrating over ${\rm d}^3\vk_d$, and further integrating over ${\rm d}^2\vell_d$ yields $\delta_D(\vtheta_d)$, which fixes $\vtheta_d=0$ after integrating over ${\rm d}^2\vtheta_d$. We can thus write
\bq
&& \T_\kappa(\vell_a,\vell_b,\vell_c,-\vell_{abc}) = \nonumber \\
&=& \left[\prod_{m=a,b,c,d} \int {\rm d}\chi_m g(\chi_m)\right] \left[\prod_{n=a,b,c} \int {\rm d}^2\vtheta_n \int \frac{{\rm d}^3\vk_n}{(2\pi)^9}e^{i\vtheta_n(\vell_n-\vk_{n,\perp}\chi_n)}e^{-ik_{n,\parallel}(\chi_n-\chi_d)}\right] \nonumber \\
&&\ \ \ \ \ \ \ \ \ \ \ \ \ \ \ \ \ \ \ \ \ \ \ \ \ \ \ \ \ \ \ \ \ \ \ \ \ \ \ \ \ \ \ \ \ \ \ \ \ \times\, T^{\rm uneq.}(\vk_a, \vk_b, \vk_c, -\vk_{abc}) \nonumber \\
&=& \left[\prod_{m=a,b,c,d}\int{\rm d}\chi_m\frac{ g(\chi_m)}{\chi_a^2\chi_b^2\chi_c^2}\right] \left[\prod_{n=a,b,c} \int \frac{{\rm d} k_{\parallel, n}}{(2\pi)^3}e^{-ik_{n, \parallel}(\chi_n-\chi_d)}\right] \nonumber \\
&& \times\, T^{\rm uneq.}\left(\left(\frac{\vell_a}{\chi_a}, k_{a,\parallel}\right), \left(\frac{\vell_b}{\chi_b}, k_{b,\parallel}\right), \left(\frac{\vell_c}{\chi_c}, k_{c,\parallel}\right), \left(-\frac{\vell_{a}}{\chi_a}-\frac{\vell_{b}}{\chi_b}-\frac{\vell_{c}}{\chi_c}, -k_{abc,\parallel}\right) \right),
\eq
where in the second equality, we have used that the integrations over ${\rm d}^2\vtheta_n$ yield $\delta_D(\vell_n - \vk_{n,\perp}\chi_n)$ Dirac-delta functions, which then fix the transverse part of the Fourier modes $\vk_n$ as $\vk_{n,\perp} = \vell_n/\chi_n$ after integrating over ${\rm d}^2\vk_{n,\perp}$. To ease the notation above, we have skipped writing the $\chi_n$ dependence on the unequal-time matter trispectrum, but added the superscript $^{\rm uneq.}$ to call attention to it. As for the case of the power spectrum, the Limber approximation amounts to setting $k_{n,\parallel}=0$ in the trispectrum arguments. Doing so, the integrals over ${\rm d}k_{n,\parallel}$ yield $\delta_D(\chi_n - \chi_d)$ Dirac-delta functions, which upon integration over ${\rm d}\chi_n$ yields
\bq
\T(\vell_a,\vell_b,\vell_c,\vell_d) &=& \int{\rm d}\chi [g(\chi)]^4 \chi^{-6} T_m\left(\frac{\vell_a}{\chi}, \frac{\vell_b}{\chi}, \frac{\vell_c}{\chi}, \frac{\vell_d}{\chi}; z(\chi)\right),
\eq
which is Eq.~(\ref{eq:limberapprox_T}).

\section{General non-Gaussian lensing covariance beyond flat-sky}\label{app:general}

In this appendix, we display and discuss the general formulae for the non-Gaussian lensing covariance without assuming the flat-sky limit nor Limber's approximation for any mode involved. Contrary to Sec.~\ref{sec:lenscovderiv_1} and the previous appendix, in which we have expanded the windowed lensing convergence in Fourier modes (cf.~Eq.~(\ref{eq:dfourierproj})), here we expand it instead in spherical harmonics $\kappa_\W(\vtheta)  = \sum_{\ell, m} a_{\ell m}Y_{\ell m}(\vtheta)$ with coefficients
\bq\label{eq:alm_kappa}
a_{\ell m} = \int {\rm d}^2\vtheta \kappa_\W(\vtheta) Y_{\ell m}^*(\vtheta) &=& \int {\rm d\chi}g(\chi)\int {\rm d}^2\vtheta \delta_{W}(\vx)Y_{\ell m}^*(\vtheta).
\eq
Being a three-dimensional quantity, we can represent $\delta_W(\vx= (\vtheta, \chi))$ in Fourier space to derive
\bq\label{eq:alm_kappa2}
a_{\ell m} &=& \int {\rm d\chi}g(\chi)\int {\rm d}^2\vtheta \int_{\vk}\tilde{\delta}_W(\vk) e^{i\vk\vx} Y_{\ell m}^*(\vtheta) \nonumber \\
&=& 4\pi i^\ell \int{\rm d}\chi g(\chi)\int_{\vk}\tilde{\delta}_W(\vk) j_\ell(k\chi)Y_{\ell m}^*(\hat{\vk}),
\eq
where we have used the plane-wave expansion of Eq.~(\ref{eq:auxeqs}) and used the orthogonality of the spherical harmonics to carry out simplifications. If $\hat{a}_{lm}$ are spherical harmonic coefficients measured from the data, then $\hat{C}(\ell) = (2\ell+1)^{-1}\sum_{m}\hat{a}_{\ell m}\hat{a}_{\ell m}^*$ is an estimator of the (monopole) angle-averaged windowed lensing power spectrum. Its expectation value can be worked out as follows
\bq\label{eq:<cl>}
\langle \hat{C}(\ell) \rangle &=& \frac{(4\pi)^2}{2\ell+1} \int {\rm d}\chi_1 {\rm d}\chi_2 g(\chi_1) g(\chi_2) \sum_m \int_{\vk_1}\int_{\vk_2} j_{\ell}(k_1\chi_1)j_{\ell}(k_2\chi_2) \nonumber \\ 
&& \times \langle\tilde{\delta}_{W}(\vk_1)\tilde{\delta}^*_{W}(\vk_2)\rangle Y_{\ell m}^*(\hat{\vk}_1) Y_{\ell m}(\hat{\vk}_2) \nonumber \\
&=& 4\pi (-1)^{\ell}\int {\rm d}\chi_1 {\rm d}\chi_2 g(\chi_1) g(\chi_2) \int_{\vk_1}\int_{\vk_2} j_{\ell}(k_1\chi_1)j_{\ell}(k_2\chi_2) \P_{\ell}(\hat{\vk_1}\hat{\vk_2}) \nonumber \\
&& \times \int_{\vp_1}\int_{\vp_2} \tilde{W}_{3D}(\vp_1)\tilde{W}_{3D}(\vp_2) (2\pi)^3\delta_D(\vk_{12} - \vp_{12}) \nonumber \\ 
&& \times P_m(|\vk_1 -\vp_1|, z_1, z_2),
\eq
where $\P_{\ell}$ is the Legendre polynomial of order $\ell$ and $z_i = z(\chi_i)$. For large $\ell$, the integrand is dominated by cases in which $k_1 \gg p_1 \lesssim 1/\chi_S$, which allows to approximate $P_m(|\vk_1 -\vp_1|) \approx P_m(k_1)$, and consequently, the integrals over the three-dimensional window function momenta immediately yield $\tilde{W}_{3D}(\vk_{12})$ (cf.~Eq.~(\ref{eq:useful1}). It is worth calling attention to the fact that, in the derivation thus far, the window function is three-dimensional, despite this being a derivation of two-dimensional lensing statistics. Here, the quantity $W(\vx)$ describes the three-dimensional geometry of the survey lightcone; both its sky-coverage, as well as its extent along the line-of-sight. For an all-sky survey (which we assume throughout this appendix for ease of discussion), this corresponds to a sphere with radius $\chi_S$. Returning to Eq.~(\ref{eq:<cl>}), we can thus write
\bq\label{eq:<cl>2}
C(\ell) \equiv \langle \hat{C}(\ell) \rangle &=& 4\pi (-1)^{\ell} \int {\rm d}\chi_1 {\rm d}\chi_2 g(\chi_1) g(\chi_2) \int_{\vk_1}\int_{\vk_2} \tilde{W}_{3D}(\vk_{12}) \nonumber \\
&& \times  j_{\ell}(k_1\chi_1)j_{\ell}(k_2\chi_2)  \P_{\ell}(\hat{\vk_1}\hat{\vk_2}) P_m(k_1, z_1, z_2).
\eq
Further noting that the integrals over $\tilde{W}_{3D}(\vk_{12})$ effectively set\footnote{We can write $\tilde{W}_{3D}(\vk_{12}) = \int {\rm d}^3\vx W_{3D}(\vx) e^{i\vk_{12}\vx}$, which is equal to $(2\pi)^3\delta_D(\vk_{12})$ if $W(\vx)=1$ everywhere. If $W(\vx)=1$ but only in a finite large volume, then one should more correctly state that $|\vk_{12}|$ should be smaller than the width of the window function in Fourier space.} $\vk_2 \approx - \vk_1$, we can write
\bq\label{eq:<cl>3}
C(\ell) \equiv \langle \hat{C}(\ell) \rangle = \frac{2}{\pi} \int {\rm d}\chi_1 {\rm d}\chi_2 g(\chi_1) g(\chi_2) \int k^2 {\rm d}k  j_{\ell}(k\chi_1)j_{\ell}(k\chi_2)  P_m(k_1, z_1, z_2).
\eq
Finally, for sufficiently large $\ell$ values (typically $\ell \gtrsim 10-20$), the spherical Bessel function terms become rapidly oscillating relative to the rest of the integrand, which allows us to write\footnote{The following orthogonality relation of the spherical Bessel functions is at the core of the argument behind the Limber approximation \cite{1953ApJ...117..134L}
\bq\label{eq:ortho1}
\int k^2 {\rm d}k j_{\ell}(k\chi_1)j_{\ell}(k\chi_2) = \frac{\pi}{2\chi_1^2} \delta_D(\chi_1 - \chi_2).
\eq
If the integrand on the left-hand side contains a function that is slowly oscillating with $k$ compared to the spherical Bessel functions (like the power spectrum is), then the above equation turns into the following approximated one
\bq\label{eq:ortho2}
\int k^2 {\rm d}k j_{\ell}(k\chi_1)j_{\ell}(k\chi_2) P_m(k) \approx  P_m\left(\frac{\nu}{\chi_1}\right)\frac{\pi}{2\chi_1^2} \delta_D(\chi_1 - \chi_2),
\eq
with $\nu = \ell + 1/2$.}
\bq\label{eq:<cl>4}
C(\ell) = \int {\rm d}\chi \left(\frac{g(\chi)}{\chi}\right)^2 P_m\left(k = \frac{\ell + 1/2}{\chi}, z(\chi)\right),
\eq
which matches Eq.~(\ref{eq:limberapprox_C}) derived assuming flat-sky.

The corresponding connected non-Gaussian covariance is formally given by (we skip looking into the Gaussian contribution for brevity, which is given by the square of Eq.~(\ref{eq:<cl>3}))
\bq\label{eq:covdeflens}
\covkappa^{\rm NG}(\ell_1, \ell_2) &=& \sum_{m_1, m_2}  \frac{\langle a_{l_1m_1} a^*_{l_1m_1} a_{l_2m_2} a^*_{l_2m_2} \rangle_c}{(2\ell_1+1)(2\ell_2+1)}.
\eq
Plugging Eq.~(\ref{eq:alm_kappa2}) in the above correlator and carrying out similar (but lengthier) algebraic steps as for the power spectrum allows us to write
\bq\label{eq:covsplit_NGapp}
\covkappa^{\rm NG}(\ell_1, \ell_2) &=&(4\pi)^2(-1)^{\ell_1+\ell_2}\left[\prod_{a = 1}^4 \int {\rm d}\chi_ag(\chi_a)\int_{\vk_a}\right]  \int_{\vp} \tilde{W}_{3D}(\vp) \tilde{W}_{3D}(\vk_{1234} - \vp) \nonumber \\
&& \times j_{\ell_1}(k_1\chi_1) j_{\ell_1}(k_2\chi_2)j_{\ell_2}(k_3\chi_3)j_{\ell_2}(k_4\chi_4)\P_{\ell_1}(\hat{\vk_1}\hat{\vk_2})\P_{\ell_2}(\hat{\vk_3}\hat{\vk_4}) \nonumber \\
&& \times T_m(\vk_1, \vk_2 + \vp, \vk_3, \vk_4 - \vp, z_1, z_2, z_3, z_4),
\eq
which tells us that the non-Gaussian lensing covariance is given by a high-dimensional integral of a generically rapidly oscillating integrand that contains the unequal-time matter trispectrum in general configurations (i.e., not restricted to $\vk_2 = -\vk_1, \vk_3 = -\vk_4$ and $z_1=z_2=z_3=z_4$ configurations). Equation (\ref{eq:covsplit_NGapp}) does not even permit a clean separation at the level of the trispectrum between super-sample contributions and the rest of the connected non-Gaussian covariance (like one could perform in the three-dimensional case in Sec.~\ref{sec:cov}).

In Sec.~\ref{sec:lenscovderiv_2}, we have displayed a derivation of the SSC lensing contribution that is valid in curved-sky cases and does not assume Limber's approximation for the long-wavelength super-survey modes (it does assume it for the sub-survey modes). On the other hand, our cNG results assume the flat-sky limit and Limber's approximation for the sub-survey modes. From Eq.~(\ref{eq:covsplit_NGapp}), the cNG contribution, in the limit where all smoothing effects due to the window are neglected, can be obtained by assuming an infinite window function, which effectively sets $\vp = 0$ and allows us to write
\bq\label{eq:covsplit_cNGapp}
\covkappa^{\rm cNG}(\ell_1, \ell_2) &=&(4\pi)^2(-1)^{\ell_1+\ell_2}\left[\prod_{a = 1}^4 \int {\rm d}\chi_ag(\chi_a)\int_{\vk_a}\right]  (2\pi)^3\delta_D(\vk_{1234}) \P_{\ell_1}(\hat{\vk_1}\hat{\vk_2})\P_{\ell_2}(\hat{\vk_3}\hat{\vk_4}) \nonumber \\
&& \times j_{\ell_1}(k_1\chi_1) j_{\ell_1}(k_2\chi_2)j_{\ell_2}(k_3\chi_3)j_{\ell_2}(k_4\chi_4)  T_m(\vk_1, \vk_2, \vk_3, \vk_4, z_1, z_2, z_3, z_4). \nonumber \\
\eq
Despite the simplification that comes from getting rid of window function momenta, the mere fact that we currently lack a general accurate description (either from theory or simulations) of $T_m(\vk_1, \vk_2, \vk_3, \vk_4, z_1, z_2, z_3, z_4)$ prevents us from tackling the problem in its full generality. Recently, Ref.~\cite{2017arXiv171107372L} has however taken interesting steps in this direction in the context of the halo model trispectrum: the many terms that contribute to the trispectrum in the halo model have certain $\vk_a$-dependencies that allow to employ Limber's approximation for some modes or mode combinations (though not all). No numerical results were shown, presumably due to the challenging numerical evaluations that still need to be performed.

%%%%%%%%%%%%%%%%%%%%%%%%%%%%%%%%%%%%%%%%%%%%%%%%%%%%%%%%%%%%%%%%%%%%%%%%%%%
\bibliography{REFS}

\end{document}